\documentclass[review]{elsarticle}

\usepackage{caption}
\usepackage{setspace}
\usepackage{amssymb}
\usepackage{multirow}
\usepackage{lscape}
\usepackage{amsthm}
\usepackage{booktabs}
\usepackage{amsmath}
\usepackage{amssymb}
\usepackage{mathrsfs}
\usepackage{longtable}
\usepackage{lscape}
\usepackage{url}
\usepackage{tabularx}
\usepackage{epsfig}
\usepackage{colortbl}
\usepackage{subfigure}
\usepackage{amsmath,amssymb,amsfonts}
\usepackage{algorithmic}
\usepackage{graphicx}
\usepackage{textcomp}
\usepackage{array}
\usepackage{epstopdf}
\usepackage{epsfig}
\usepackage{graphicx}
\usepackage{stfloats}
\usepackage{amsmath,amssymb,amsfonts}
\usepackage{algorithmic}
\usepackage{textcomp}
\usepackage{diagbox}
\usepackage{float}
\usepackage{subfigure}
\usepackage{color}

\newtheorem{definition}{Definition}[section]
\usepackage{lineno,hyperref}
\modulolinenumbers[5]

\begin{document}

\begin{frontmatter}



\title{Identifying influential nodes in complex networks:  Effective distance gravity model}

\author[address1,address2]{Qiuyan Shang}
\author[address1]{Yong Deng \corref{label1}}
\author[address3,address4,]{Kang Hao Cheong\corref{label1}}
\ead{kanghao\_cheong@sutd.edu.sg}
\address[address1]{Institute of Fundamental and Frontier Science, University of Electronic Science and Technology of China, Chengdu, 610054, China}
\address[address2]{Yingcai Honors of school, University of Electronic Science and Technology of China, Chengdu, 610054, China}
\address[address3]{Science and Math Cluster, Singapore University of Technology and Design (SUTD), S487372, Singapore}
\address[address4]{SUTD-Massachusetts Institute of Technology International Design Centre, Singapore}
\cortext[label1]{Corresponding author at: Institute of Fundamental and Frontier Science, University of Electronic Science and Technology of China, Chengdu, 610054, China. E-mail: dengentropy@uestc.edu.cn, prof.deng@hotmail.com. (Yong Deng)}

\begin{abstract}

The identification of important nodes in complex networks is an area of exciting growth due to its applications across various disciplines like disease controlling, community finding, data mining, network system controlling, just to name a few. Many measures have thus been proposed to date, and these measures are either based on the locality of nodes or the global nature of the network. These measures typically use distance based on the concept of traditional Euclidean Distance, which only focus on the local static geographic distance between nodes but ignore the interaction between nodes in real networks. However, a variety of factors should be considered for the purpose of identifying influential nodes, such as degree, edge, direction and weight. Some methods based on evidence theory have also been proposed. In this paper, we have proposed an original and novel gravity model with effective distance for identifying influential nodes based on information fusion and multi-level processing. Our method is able to comprehensively consider the global and local information of the complex network, and also utilize the effective distance to replace the Euclidean Distance. This allows us to fully consider the complex topological structure of the real-world network, as well as the dynamic interaction information between nodes. In order to validate the effectiveness of our proposed method, we have utilized the susceptible infected (SI) model to carry out a variety of simulations on eight different real-world networks using six existing well-known methods. The experimental results indicate the reasonableness and effectiveness of our proposed method.
\end{abstract}

\end{frontmatter}

\section{Introduction}
In recent years, the study of complex network\cite{xie2019high,de2020efficient} has attracted immense attention\cite{gao2019repulsive}. Many real-world problems\cite{zareie2019identification,weibo2019network1} can be analyzed as part of network science\cite{gosak2018network,gallos2012small} for further research\cite{wang2018exploiting}, such as Internet security, network control system\cite{zhu2014maximizing,Weibo2019network2} and social network. Hence, the identification of influential nodes in complex networks play an important role\cite{sun2018identifying} in both structural and functional aspects\cite{li2019identifying,wentao2018information}, and is an important area of research\cite{chen2012identifying}. The identification of influential nodes can be applied across various fields\cite{li2018evidential} such as disease\cite{wang2016statistical}, network system\cite{faris2019intelligent}, biology\cite{majhi2019chimera}, social system\cite{gallos2012people,zareie2018influence,zareie2019identification}, time series\cite{pravilovic2017using}, information propagation\cite{xu2019identifying} and Parrondo's paradox\cite{tan2017nomadic,tan2019predator,cheong2018time,cheong2016paradoxical,cheong2018multicellular}. Besides, identifying the vital nodes\cite{Zareie2018hierarchical} can allow us to discover and address real-world problems\cite{helbing2015saving,pei2017efficient} such as transportation hubs identifying, influence maximizing, rumor controlling\cite{hosni2020minimizing}, disease controlling\cite{kitsak2010identification}, advertising and community finding\cite{zheng2017finding,he2018hidden}.

  Many methods have been proposed to assess the influence of nodes\cite{wu2018power,saito2016super}. These methods can be classified under two broad categories: locality of nodes and the global nature of the network. One view is that the influence of nodes often depends on its neighbors, such as degree centrality (DC)\cite{chen2012identifying}, K-shell decomposition method (KS)\cite{kitsak2010identification}, semi-local centrality\cite{chen2012identifying} and PageRank\cite{lu2016vital}. For DC, the influence of nodes is determined by the number of neighbors; a node with many neighbors is of high influence. The KS suggests that the influence of nodes is related to their topological properties in the local area. The more central the node is in the local structure, the more influential the node is. PageRank, the algorithm for random walking\cite{masuda2017random} states that the influence of nodes is not only dependent on the number of neighbors, but also related to the quality of neighbors\cite{li2019identifying}. It works by simulating the process of browsing the web. In general, it has good performance on directed network, but does not perform very well on undirected network. 
  
  Another view is that the influence of nodes mainly depends on paths in the network. For example, closeness centrality (CC)\cite{wang2019modified} and betweenness centrality (BC)\cite{li2019identifying} are representatives of such algorithm. CC suggests that the shorter the average distance between a node and other nodes, that is, the closer the node is located to the center, the more influential the node is. However, BC claims that the influence of a node is mainly determined by the number of shortest paths through it.  Although BC and CC algorithms can often give better results than other algorithms, they are very sensitive to network structure\cite{wen2020Vital} and the complexity of these algorithms is high with many limitations as well. The above algorithms are either neighborhood-based local method, or the path-based global method. 

Recently, inspired by the law of gravity, Li et al. proposed an algorithm based on the gravity model, called gravity model (GM)\cite{li2019identifying}. Liu et al. further proposed a more generalized weighted gravity model, called generalized mechanics model (GMM)\cite{Liu2020GMMKBS}. The two models take into account both the neighborhood-based local information and the path-based global information. They are also applicable on both directed and undirected networks, and have proven to be effective and feasible. However, the distance in the algorithm based on the gravity model mainly utilizes the traditional Euclidean Distance\cite{danielsson1980euclidean}, focusing only on the local static geographical location between nodes, while ignoring the interaction between nodes in the actual network. In order to address this critical gap, we propose an original and novel method called the effective distance gravity model. On the basis of GM, the original Euclidean Distance\cite{brockmann2013hidden} is now replaced by the effective distance proposed by Brockmann et al. Effective distance is an abstract concept of distance derived from the idea of probability. It mainly pays attention to the interaction of nodes in the network and uses it as the main basis for judging. The core of effective distance is to discover the most probable path between two points by calculating the probability through the adjacency matrix. The effective distance fully considers the potential dynamic information interaction between nodes in the actual network. Therefore, our proposed effective distance gravity model takes into account not only the network local and global information, but also the potential dynamic interaction between nodes, such that a node with more neighbors and shorter effective distance between the other nodes is more influential. Based on our proposed method, we have carried out a variety of experiments on eight real networks using the susceptible infected (SI) model\cite{meng2014adaptive}, and compared it with six existing well-known identification methods. Our experimental results indicate the robustness and reasonableness of our proposed method over existing methods. The paper is organized as follows. In Section 2, we describe the parameters used in this paper, an overview of several well-known node identification measures is given. The concept of effective distance will also be introduced. In Section 3, a new identification of influential nodes measure: effective distance gravity model is proposed. In Section 4, a variety of experiments and comparisons with other measures are then illustrated to show the feasibility and effectiveness of our proposed method. We conclude our study in Section 5.

\section{Preliminaries}
In an undirected graph $G=(V,E)$,where the $V$ represents the set of nodes and $E$ represents the set of links\cite{le2020mining}. And the number of nodes in the graph is denoted as $n$, where $n=|V|$. The adjacency matrix of graph $G$ is $A =\left\{ a_{i j}\right\}$, where $a_{i j}=1$ if there is an edge between node $i$ and node $j$.
\subsection{Centrality measures}
\begin{definition}\label{b1}
 $Degree \ centrality (DC)$ identifies the importance of a node by comparing degree of the node. The node with large degree is of high influence\cite{li2020key}. $DC(i)$ of each node $i$ can be obtained by the following formula.
\begin{equation}\label{q1}
DC(i)=\sum_{j}^{N} a_{i j}=k_i
\end{equation}
Where $k_i$ is the degree of node $i$\cite{ding2019consensus}.

\end{definition}

\begin{definition}\label{b2}
The definition of $Betweenness \ centrality (BC)$\cite{Liu2020GMMKBS} is as follows. BC measures the importance of a node by the number of shortest paths through it. The more the number of shortest paths through node i, the more important node i is in the network.
\begin{equation}\label{q2}
BC(i)=\sum_{j, k \ne i} \frac{N_{j k}(i)}{N_{j k}}
\end{equation}
Where $N_{j k}$ represents the number of shortest paths from node $j$ to node $k$, and $N_{jk}(i)$ is the number of $N_{jk}$ through node $i$.
\end{definition}

\begin{definition}\label{b3}
$Closeness \ centrality (CC)$\cite{wang2019modified} evaluates the influence of nodes by the reciprocal of the sum of shortest path between nodes. The higher the $CC(i)$ is, the more important the node $i$ is. 
\begin{equation}\label{q3}
CC(i)=\frac{1}{\sum_{j}^{N} d_{i j}}
\end{equation}
where $d_{ij}$ denotes the length of shortest path between node $i$ and node $j$.\end{definition}

\begin{definition}\label{b4}
$Eigenvector \ centrality (EC)$\cite{Liu2020GMMKBS} is a complex method, which claims the influence of a node is determined not only by the number of neighbors, but also by the importance of the them. $EC(i)$, the centrality scores of node $i$, can be calculated by the formula below.
\begin{equation}\label{q4}
EC(i)=\frac{1}{\lambda}\sum_{j=1}^n (a_{ij}x_j)
\end{equation}
The largest eigenvalue of $A$ is be represanted by $\lambda$ and $x_j$ is the value of $jth$ entry of the eigenvector corresponding to $\lambda$.
\end{definition}

\begin{definition}\label{b5}
$PageRank (PC)$ uses an iterative approach to obtain the influence of nodes, and it is very effective to calculate the importance of nodes in the directed network. $PC(i)$ of node $i$\cite{lu2016vital} can be obtained by following formula.
\begin{equation}\label{q5}
PC(i)^q=\sum_{j=1}^n (a_{ij}\frac{PC(j)^{q-1}}{k_j})
\end{equation}
The influence score of node $i$ in step $q$ is denoted as $PC(i)^q$. The higher the $PC$ score when the PC finally converges is, the more vital the node is.
\end{definition}

\subsection{Gravity model (GM)}
 The $GM$ is defined by the gravity formula. The influence of a node can be estimated by GM\cite{li2019identifying} as follows. 
  \begin{equation}\label{q6}
C(i)=\sum_{i \ne j}\frac{k_i\times k_j}{(d_{ij})^2}
\end{equation}
 $C(i)$ represents the centrality score of node $i$,the degree of the node $j$ is denoted as $k_j$. The shortest path between node $i$ and node $j$ can be represented as $d_{i j}$. In particular, the distance here uses Euclidean Distance.

\subsection{Effective distance (ED)}
  Effective distance\cite{brockmann2013hidden} is an abstract distance based on probability, which mainly focuses on the potential information interaction between nodes in the real complex networks. The definition of $ED$ is as follows.
\begin{equation}\label{q7}
D_{m n}= \min{\left\{1-log_2(P_{mn}^*)\right\}}
\end{equation}
Where $D_{m n}$ is the value of effective distance from node $m$ to node $n$ and $P_{m n}$ is the probability from node $m$ to node $n$, which can also be obtained by the product of multiple probabilities in the graph. For example $P_{m n}^*$  can be calculated by $P_{m n}^* = P_{m l}\times P_{l s}\times ...\times P_{k n}$ , which is similar to Markov Chain.
The $P_{m n}$ is calculated as follows.
 \begin{equation}\label{q8}
P_{m n}=\frac{a_{mn}}{k_m}      (m \ne n)
\end{equation}
Where $k_m$ is the degree of node $m$, $a_{m n}$ is the element in the adjacency matrix of graph $G$.

\section{Proposed method}\label{s3}

\subsection{Effective distance gravity model (EffG)}
In reality, many problems can be analyzed as part of network science for further research. The structure and function of these actual networks are often more complicated than we think. Thus, for the identification of influential nodes, only use neighborhood-based local properties of the network but ignore the global connectivity are unadvisable. Similarly, it is not feasible to only consider the path-based globality of the network but ignore the local properties of the node. Individually speaking, the two properties should be considered together in order to achieve a good effect. Therefore, we consider a comprehensive consideration of the degree of nodes and the path of the network.

At the same time, considering the complex structure and the evolution of the complex network, the structure of network is likely to be unstable. Therefore, there may be some errors for the node identification. However, multiple cumulative summing can effectively solve this problem.

In addition, since the distance of most of the measures is conventional Euclidean Distance, which is the static geographic distance between two points. However, considering the multiple complexity fusion of complex networks, we cannot simply think that the true structure of the network is the exhibited geometry structure currently. The potential interaction of information and energy between nodes may lead to changes in the network structure. Therefore, we believe that complex networks are likely to have a potential geometric structure, which often drives many dynamics propagation processes, such as the spread of epidemics and rumors. In this situation, it is obviously inadequate to use a simple static geometric metric such as Euclidean distance. Dirk Brockmann and Dirk Helbing claimed that if probability is used to construct a new distance metric to replace conventional geographic distance, then the complex space-time patterns can be reduced to surprisingly simple, homogeneous wave propagation patterns. Their experimental results showed that the effective distance can reliably predict the arrival time of the disease. Thus, we believe that the effective distance fully takes into account a potential topology of complex network due to the dynamic information interaction between nodes\cite{brockmann2013hidden}, which has certain significance for the identification of influential node in the complex network. Consequently, we consider replacing the conventional Euclidean Distance with the effective distance proposed by Brockmann to further optimize the algorithm.

In summary, we proposed an effective distance gravity model, which not only comprehensively considers the local and global network structural indicators, but also reduces the identification error caused by the unstable structure of the complex network through cumulative summation. At the same time, by replacing Euclidean Distance with effective distance, the dynamic interaction between nodes and the potential complex topology of the network are fully considered.
Therefore, the influence of the node can be estimated as
\begin{equation}	
C_{EffG}(i) = \sum_{i \ne j}\frac{k_i \times k_j}{D_{ij}^2}		  
\end{equation}
 Where $k_i$ and $k_j$ are the degree of node $i$ and node $j$, $D_{ij}$ is the effective distance from node $i$ to node $j$. And $C_{EffG}(i)$ represents the centrality scores of node $i$. The whole steps and calculation process of proposed method are shown in Fig.1.
 \begin{figure*}[!htbp]
\centering
\includegraphics[width=0.72\textwidth]{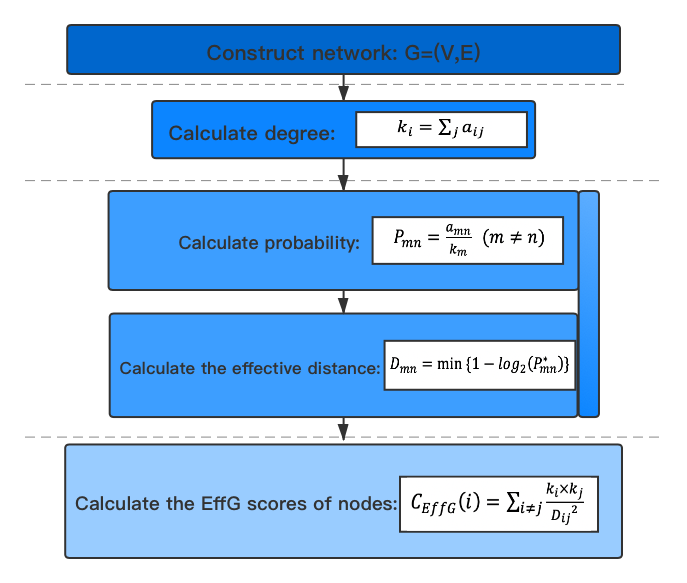}\\
\caption{\textbf{The flow chart of our proposed method.} The first step is to build a network.The second step is to calculate the degree of each node. The third and fourth steps are combined to obtain the effective distance between the nodes. Finally, the fifth step calculates the centrality scores of each node.}
\end{figure*}

\subsection{An example}
 In order to better explain our proposed identification method EffG, here a simple example is given to help understand how EffG works in the network. We take $node 2$ as an example to calculate the EffG scores of it.
 The Fig.2(a) is the graph of a network, which adjacency matrix is the Fig.2(b).
 \begin{figure}[!htbp]
 \centering
 \subfigure[The graph of a simple network]{
 \includegraphics[width=0.3\textwidth]{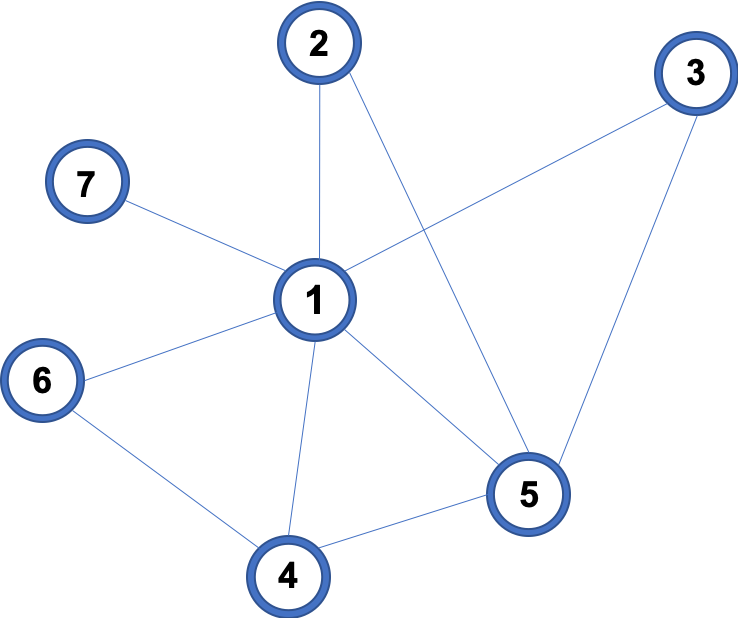}}%
  \quad
 \subfigure[The adjadency matrix of network]{
 \includegraphics[width=0.3\textwidth]{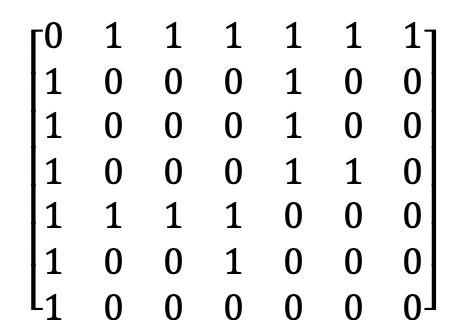}}%
 \caption{A simple network with seven nodes}
  \end{figure}
 
  The degree of each node is presented in the Table \uppercase\expandafter{\romannumeral1}. 
  \begin{table}[!htbp]
  \centering
  \caption{The degree of each node in Fig.2.}
  \begin{tabular}{cccccccc}
  \hline
  Node &$node1$& $node2$& $node3$& $node4$& $node5$& $node6$& $node7$ \\ 
  \hline
  degree& 6& 2& 2& 3& 4& 2& 1 \\ 
  \hline
 \end{tabular}
 \end{table}
  
  First, we calculate the effective distance between nodes by the Function (7). In particularly, Under normal circumstances , the $P_{i j} \ne P_{j i}$ and $D_{i j} \ne D_{j i}$. Besides, because $P_{i i}$, the probability from node $i$ to itself is often zero, the distance from node $i$ to itself which denoted as $D_{i i}$ is also infinite.\\
The effective distance between $node 2$ and $node 7$ is calculated as:
\begin{equation}
\begin{aligned}
D_{27}&=\min{\left\{1- \log_2(P_{27}^*)\right\}}\\
&= 1- \log_2(\max{\left\{P_{27}^*\right\}})\\
&= 1- \log_2(\max{\left\{P_{27},P_{21} \times P_{17}, P_{25} \times P_{51} \times P_{57},...\right\}})\\
&=1-\log_2(P_{21} \times P_{17}) \\
&=1-\log_2(\frac{1}{2} \times \frac{1}{6}) \\
&=4.5850\\
D_{72} &=\min{\left\{1- \log_2(P_{72}^*)\right\}}\\
&=1-\log2(\max{\left\{P_{72}^*\right\}})\\
&=1-\log_2(P_{71} \times P_{12}) \\
&=1-\log_2(\frac{1}{1} \times \frac{1}{6})\\
&=3.5850
\end{aligned}
\end{equation}
\nonumber
\indent Using the same method of $D_{27}$ and $D_{72}$, the effective distance between other nodes and $node 2$ can be calculated. The result is in the Table \uppercase\expandafter{\romannumeral2}.\\
  \begin{table}
  \centering
  \caption{Effective Distance (ED) between nodes in Fig.2.}
  \begin{tabular}{cccccccc}
  \hline
   $D_{ij}$& $D_{21}$& $D_{22}$& $D_{23}$& $D_{24}$& $D_{25}$& $D_{26}$& $D_{27}$\\ 
  \hline
  ED& 2.0000& $+ \infty $& 4.0000 &4.0000& 2.0000& 4.5850& 4.5850 \\ 
  \hline
 \end{tabular}
 \end{table}
\indent Then, the EffG scores of node 2 can be calculated as follows:\\
\begin{center}
$C_{EffG}(2) = \sum_{j \ne 2}\frac{k_2 \times k_j}{D_{2 j}^2} = 5.9104 $\\
\end{center}

\indent The EffG scores of the other nodes can be calculated by the same method, which are showed as follows:\\
\begin{center}
$C_{EffG}(1) = \sum_{j \ne 1}\frac{k_1 \times k_j}{D_{1 j}^2} = 6.5358$\\
$C_{EffG}(3) = \sum_{j \ne 3}\frac{k_3 \times k_j}{D_{3 j}^2} = 5.9104$\\
$C_{EffG}(4) = \sum_{j \ne 4}\frac{k_4 \times k_j}{D_{4 j}^2} = 6.0704$\\
$C_{EffG}(5) = \sum_{j \ne 5}\frac{k_5 \times k_j}{D_{5 j}^2} = 6.2865$\\
$C_{EffG}(6) = \sum_{j \ne 6}\frac{k_6 \times k_j}{D_{6 j}^2} = 5.5981 $\\
$C_{EffG}(7) = \sum_{j \ne 7}\frac{k_7 \times k_j}{D_{7 j}^2} = 1.0115 $\\
\end{center}

\indent From the Fig2.(a). we can see that the $node 1$ is the central of the network which has the most strong connection with others and covers the most shortest paths in the network. Without $node 1$, the network will be broken into multiple isolated parts. Thus, it is reasonable that the $node 1$ is the most influential node in this network. Besides, the $node 7$ can be seen that is the least influential in the network, and the EffG score that matches it is the lowest. And the importance of $node 2$ and $node 3$ in this network is basically the same, they also have the same EffG score. This simple example shows that our proposed method EffG is practical and objective.

\section{Application}\label{s4}
To verify the feasibility and effectiveness of our proposed method, six experiments were performed on eight actual networks, in comparison with six existing well-known methods.
\subsection{Datasets}
The experiment was conducted on eight real-world networks, Jazz, NS\cite{wentao2019nodes}, GrQc, Email, EEC, Facebook\cite{Liu2020GMMKBS}, PB\cite{wen2020vulnerability} and USAir\cite{li2019identifying}, including two communication networks (Email, EEC), one transportation network (USAir), two social networks (Facebook, PB) and three cooperative networks (Jazz, NS, GrQc). Among them, Email is a network where users send emails and communicate with each other. EEC is a network where European researchers do exchange of emails. USAir represents a US air transportation network. Facebook describes a social network derived from Facebook. PB is a blog network. Jazz describes a practical network of Jazz musician collaborations. NS is a network where scientists collaborate and work together. GrQc is a network published on preprint. Other relevant information about the network is displayed on Table \uppercase\expandafter{\romannumeral3}.
  \begin{table}[!htbp]
  \centering
  \caption{\textbf{The basic topology information of the eight actual networks.} n and m are the number of nodes and edges of the network, $<k>$ and $<d>$ are the average degree and average distance of the network. C and r are the network's clustering coefficient\cite{wentao2018evaluating} and assortative coefficient\cite{wentao2019structure}.}
  \begin{tabular}{ ccccccc}
  \hline
  Networks& n& m& $<k>$& $<d>$& C& r\\ 
  \hline
  Jazz& 198& 2472& 27.6970& 2.2350& 0.6334& 0.0202 \\ 
  NS& 379& 914& 4.4832& 6.0419& 0.7981& -0.0817\\
  GrQc& 4158& 13422& 6.4560& 6.0494& 0.6648& 0.6392\\
  EEC& 986& 16064& 32.5842& 2.5869& 0.4505& -0.0257\\
  Email& 1133& 5451& 9.6222& 3.7160& 0.1101& 0.0782\\
  PB& 1222& 16714& 27.3553 &2.7375& 0.3600& -0.2213\\
  Facebook& 4039& 88234& 43.6910& 3.6925& 0.6170& 0.0636\\
  USAir& 332& 2126& 12.8072& 2.7381& 0.7494& -0.2079\\
  \hline
 \end{tabular}
 \end{table}
\subsection{Centrality scores of nodes}
In this experiment, our proposed method (EffG)was used to calculate centrality scores in six real-world networks we provided. Five existing well-known methods (DC, CC, BC, PC, Gravity) were used in the same networks for comparison. The experimental results are shown in Figs.3-8. The importance of the node is reflected by the color of the node in the heat map. The darker the color of node is, the more influential the node is.\\
\indent As can be seen, the distribution of relative importance of nodes is basically consistent, although the centrality scores calculated by CC and EffG are higher and the value of BC is lower. Besides, it can be easily found in figs.4-8. that the centrality scores calculated by BC, DC and PC are difficult to distinguish in the figures because the scores are all quite low. However, the centrality scores calculated by CC, Gravity model and EffG are well distinguished, especially CC and EffG. Moreover, the distribution of relative importance of nodes in CC and EffG is more similar. Consequently, our proposed method, EffG, can be found to be convenient for us to distinguish the relative importance of nodes with high accuracy.
\begin{figure}[!htbp]
\centering
\includegraphics[width=0.8\textwidth]{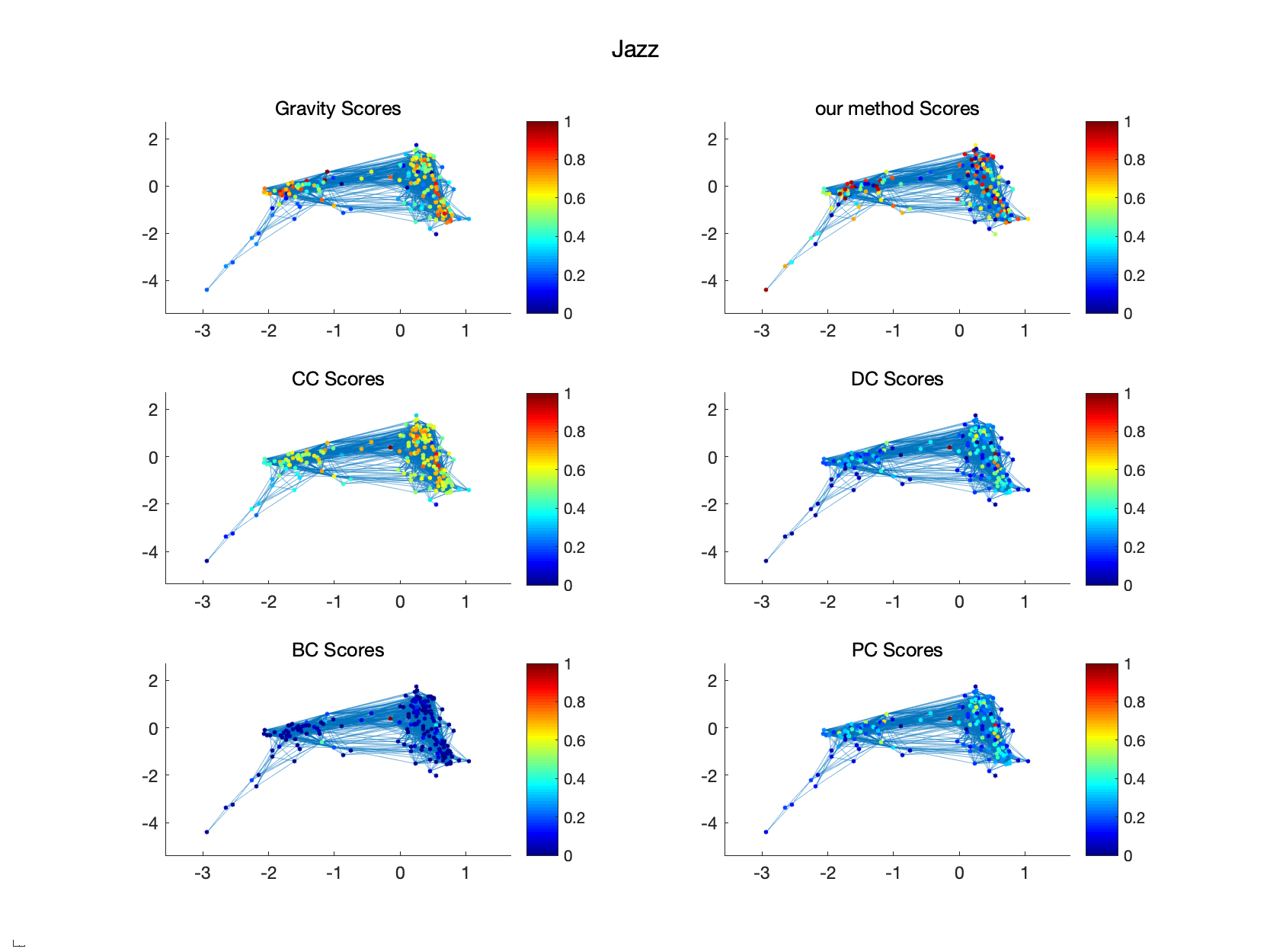}
\caption{This figure compares the centrality scores of different measures in Jazz. The distribution of Gravity model, CC and our proposed method is similar.}
\end{figure}
\begin{figure}[!htbp]
\centering
\includegraphics[width=0.8\textwidth]{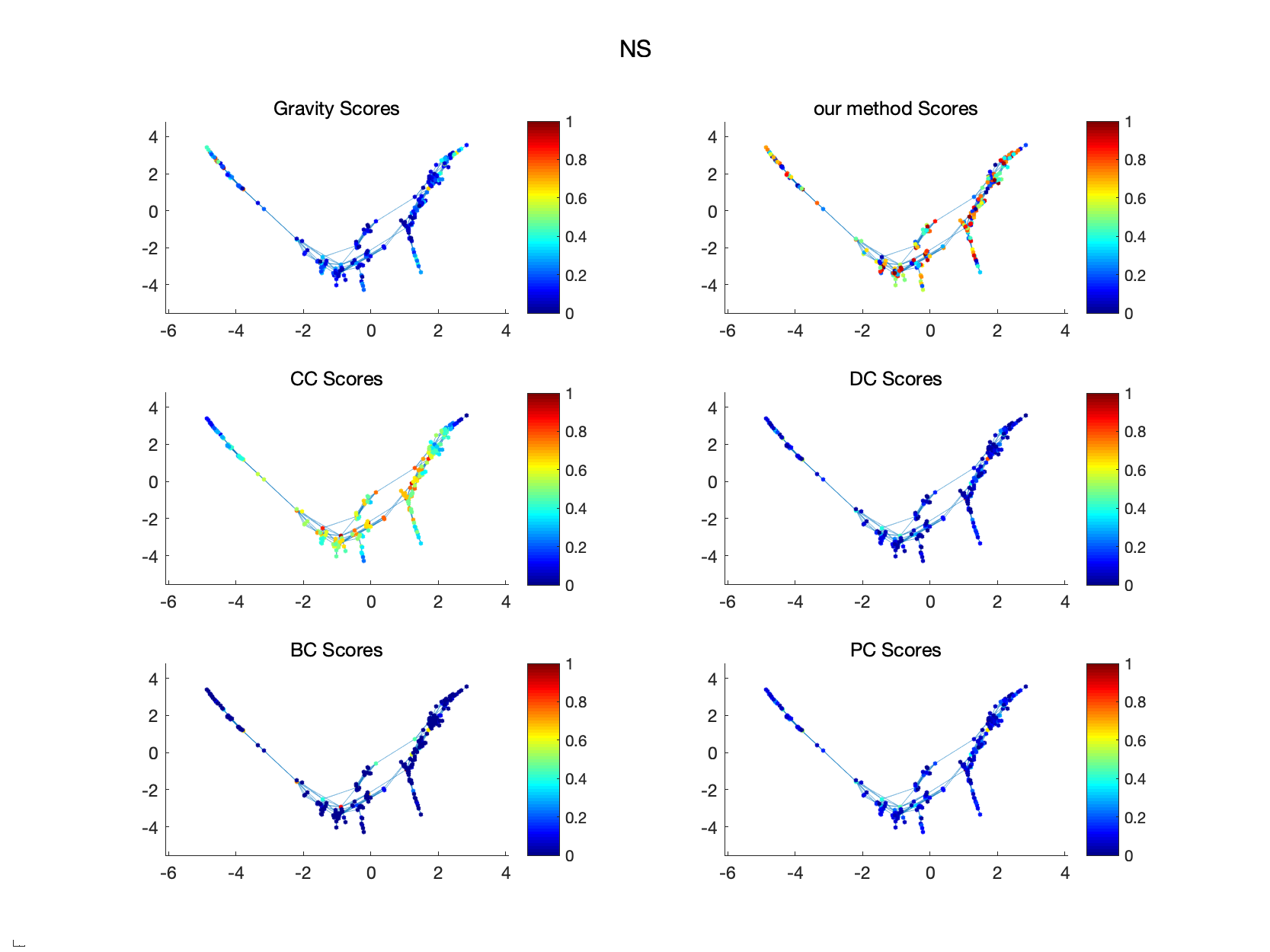}
\caption{This figure compares the centrality scores of different measures in NS. DC, Gravity, BC and PC are basically the same. Our method and CC are similar in the distribution of cital node.}
\end{figure}
\begin{figure}[!htbp]
\centering
\includegraphics[width=0.8\textwidth]{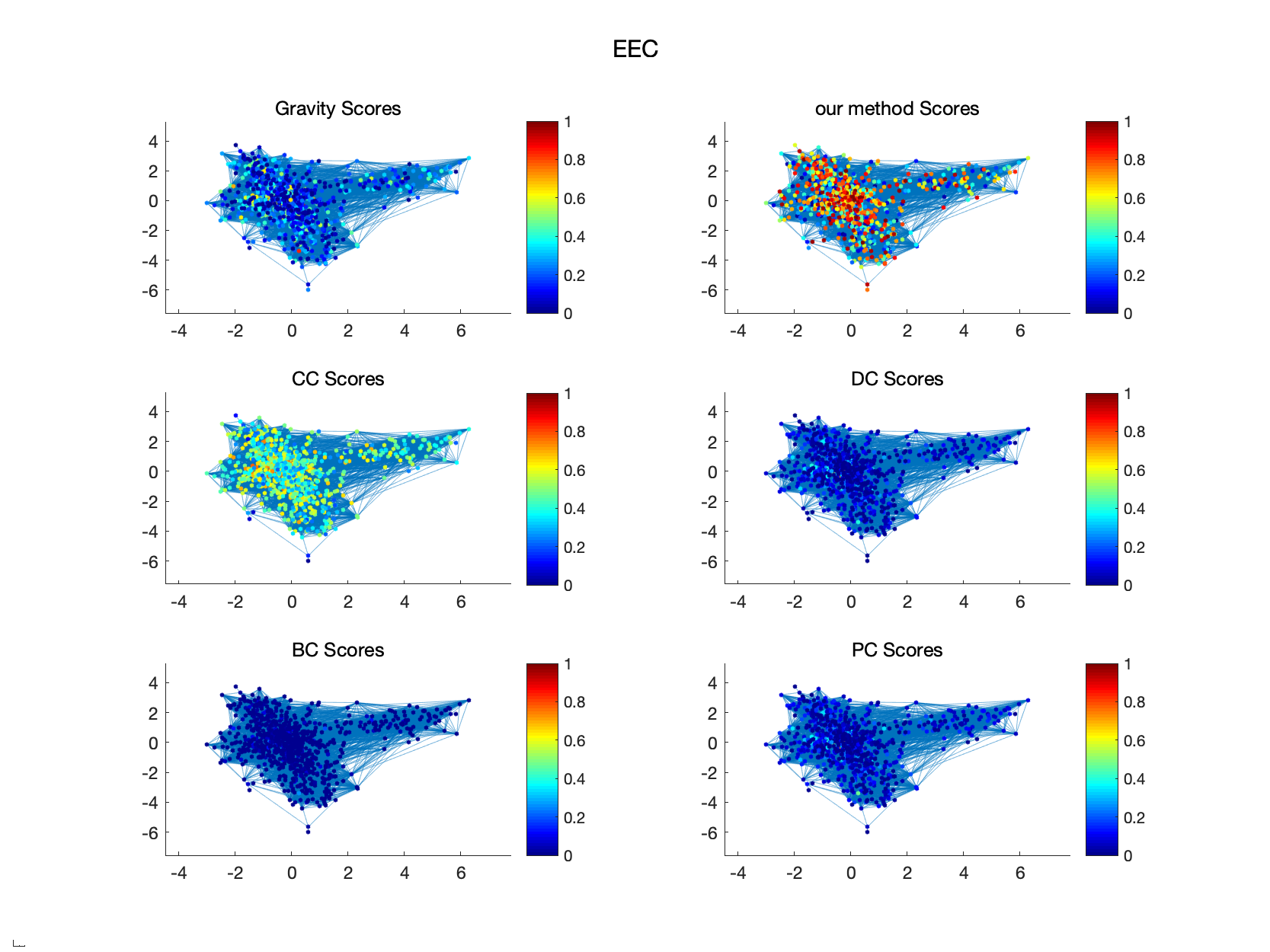}
\caption{This figure compares the centrality scores of different measures in EEC. The value of proposed method and CC is higher, and BC is difficult to distinguish the different nodes.}
\end{figure}

\begin{figure}[!htbp]
\centering
\includegraphics[width=0.8\textwidth]{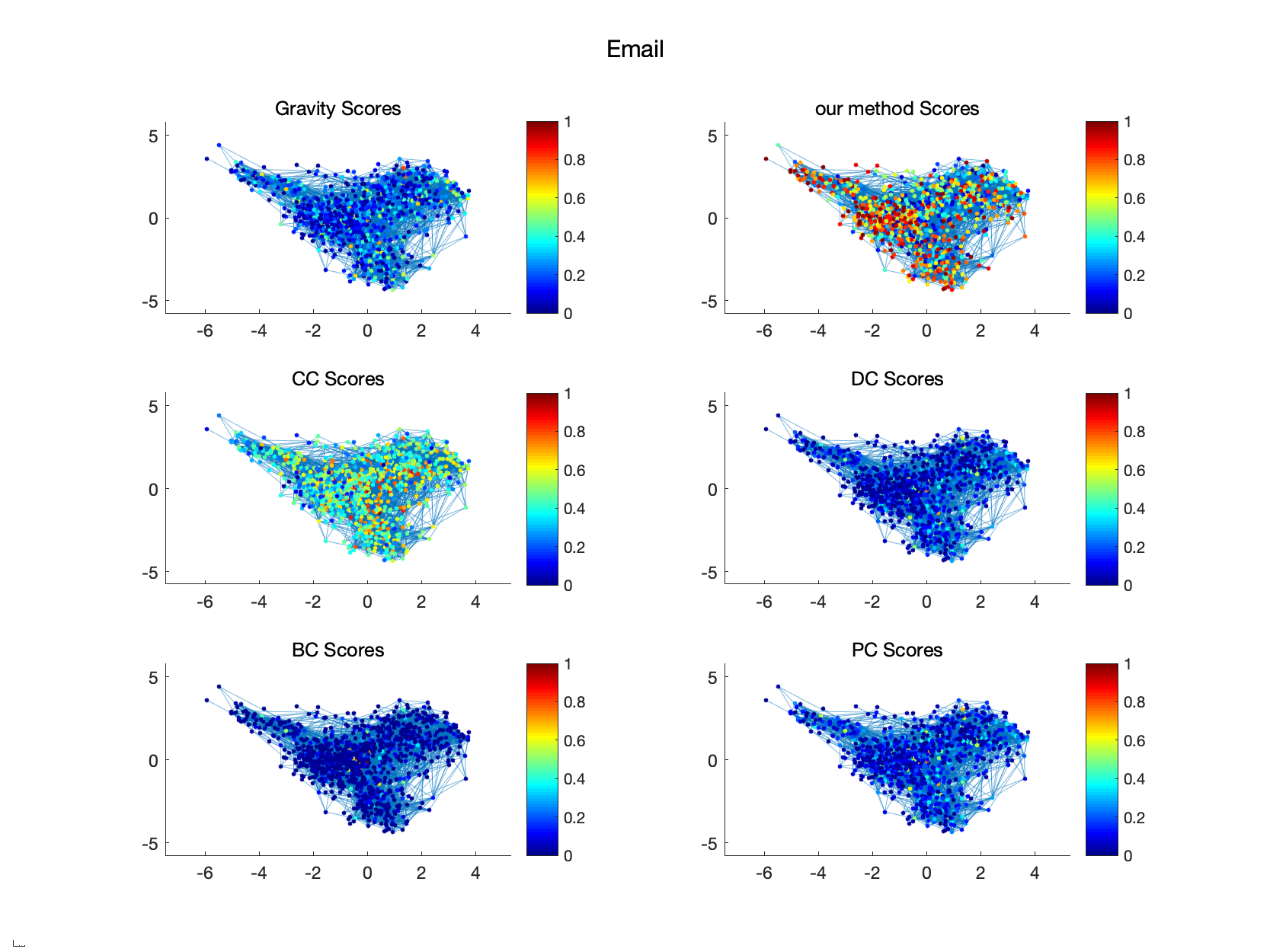}
\caption{This figure compares the centrality scores of different measures in Email. DC and PC is almost the same, and the distribution of CC and proposed method is similar although the value of our method is higher.}
\end{figure}

\begin{figure}[!htbp]
\centering
\includegraphics[width=0.8\textwidth]{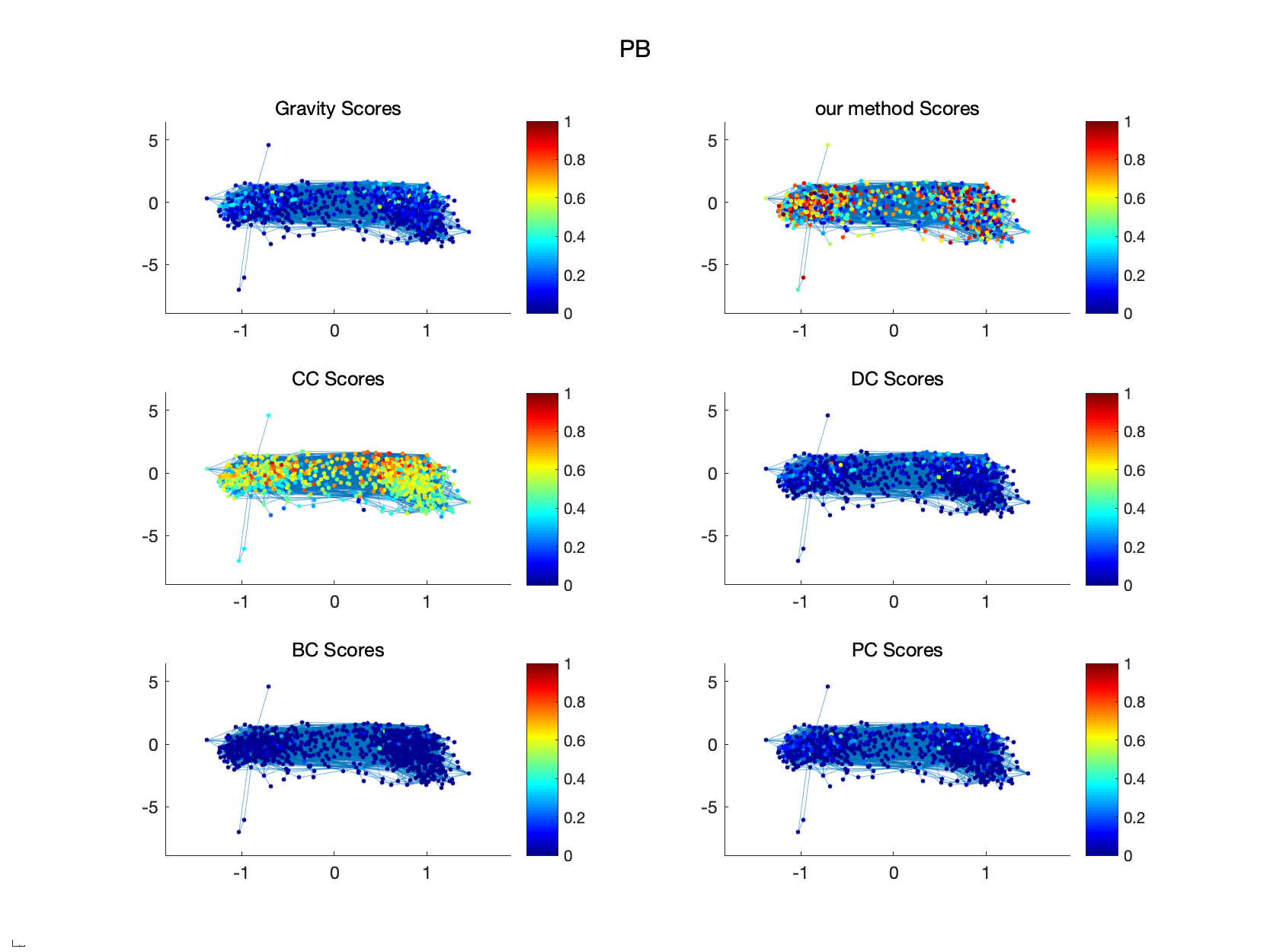}
\caption{This figure compares the centrality scores of different measures in PB. The value of CC is highest, the value of BC is lowest. The distribution of our method is little different from the others except for CC.}
\end{figure}
\begin{figure}[!htbp]
\centering
\includegraphics[width=0.8\textwidth]{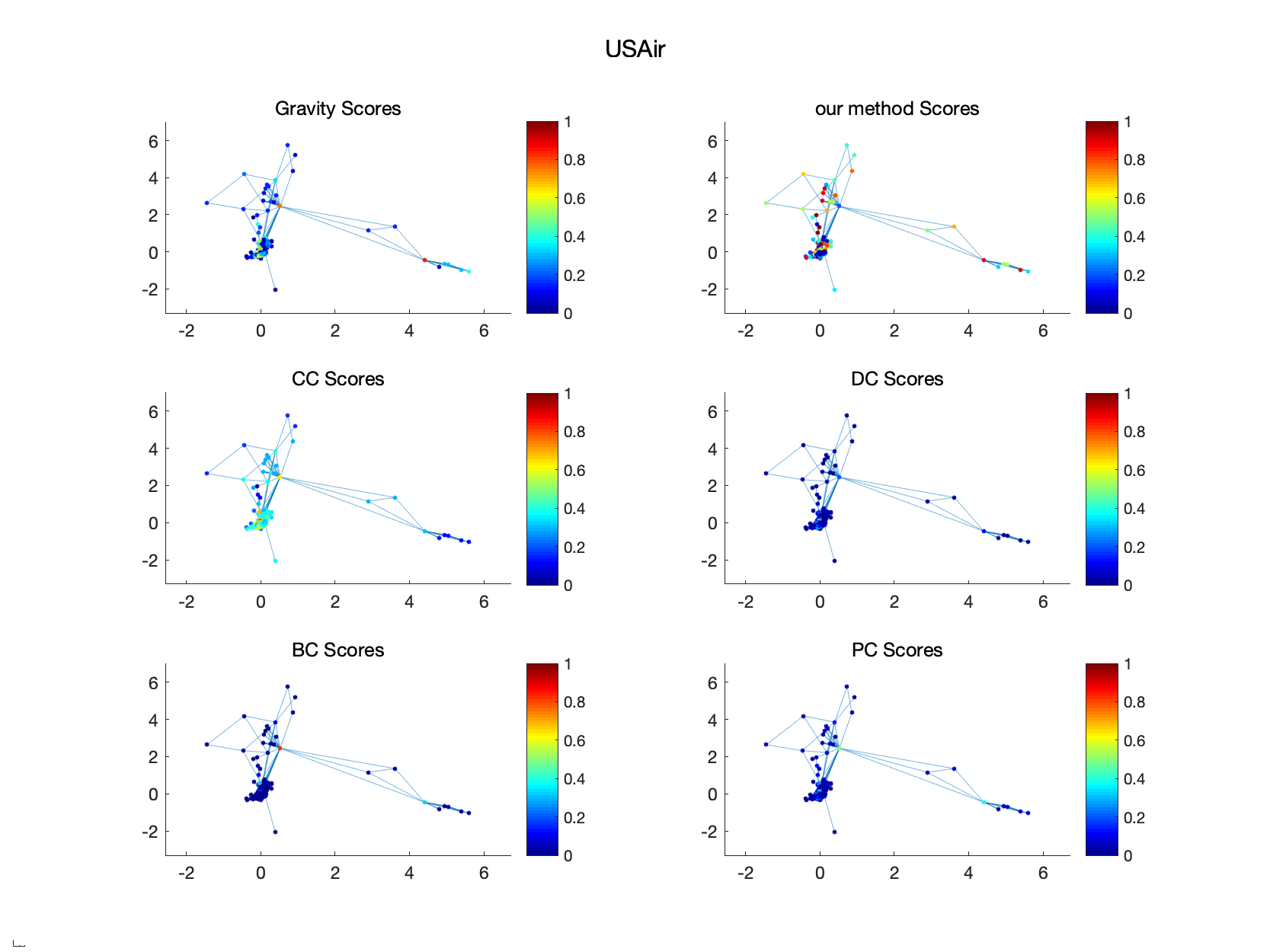}
\caption{This figure compares the centrality scores of different measures in USAir. DC and PC is basically the same, and CC can clearly be seen the difference in global influence of nodes.}
\end{figure}

\subsection{Evaluating with susceptible infected (SI) model }
The susceptible infected (SI) model\cite{meng2014adaptive} can be used to estimate the node's capability of transmission in the network, which indirectly reflects the influence of the node. In the SI model, there are two compartments deserve our attention: (1) susceptible state (2) infection state. In the process, the infected nodes infect the surrounding susceptible nodes with a certain probability. The parameters utilized in the SI model are t, F (t), $\beta$ and N. The experimental simulation time of the susceptible infected model is denoted by t. $\beta$ represents the probability of nodes infection. N is the number of experiments. The average number of infected nodes at time t, denoted by F(t). It can be easily understood that the more important the node has the greater the influence. And under the condition that the infection time t and the infection probability B are both the same, the more influential node will cause more surrounding nodes to be infected. Hence, F(t) reflects the influence of the initial infected node. The node with higher F(t) is of greater importance.\\
\indent In order to estimate the capability of different measures in identifying the vital nodes, the SI model was applied on eight different real-world networks. In the experiments, the top-100 nodes ranked by different methods was selected firstly. After that, the top-100 nodes were used as the initial infection nodes in the SI model separately. Finally, the average number of infected nodes F(t) was calculated for each method respectively. In particular, the SI model in experiments is given the same propagation probability$\beta$ to control the variables, and the $\beta$ was set to be 0.2 in our experiment.\\
\indent The experimental results are shown in Figs.9-16. The node with more final infected nodes is of greater importance. Hence, the faster the curve rises and the higher the curve is, the more influential the nodes in the initial infection set are. That is to say, the more effective the identification method is. 

\begin{figure}[!htbp]
\centering
\includegraphics[width=0.8\textwidth]{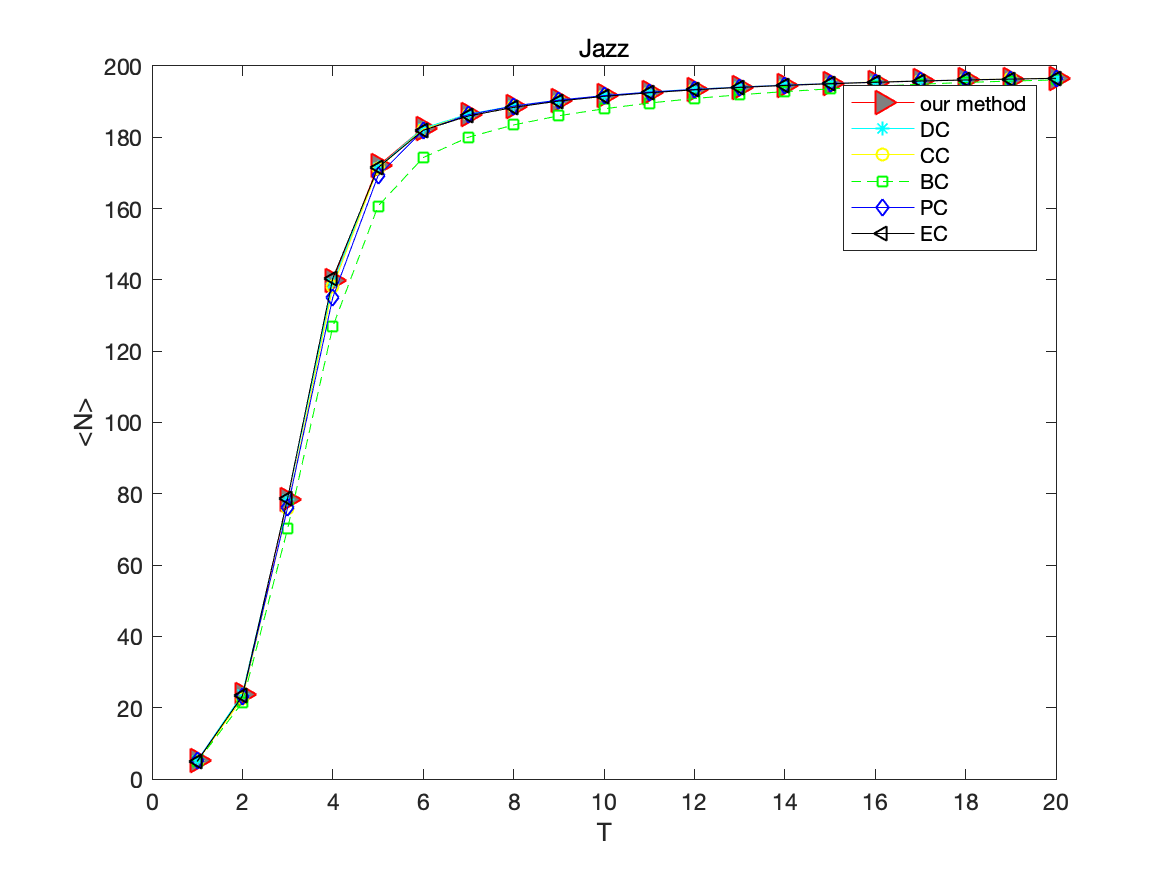}
\caption{The figure compares the infection ability of the top-100 nodes selected by different methods in Jazz. All methods except for BC are basically the same and our method is the highest and the curve rises faster than others.}
\end{figure}
\begin{figure}[!htbp]
\centering
\includegraphics[width=0.8\textwidth]{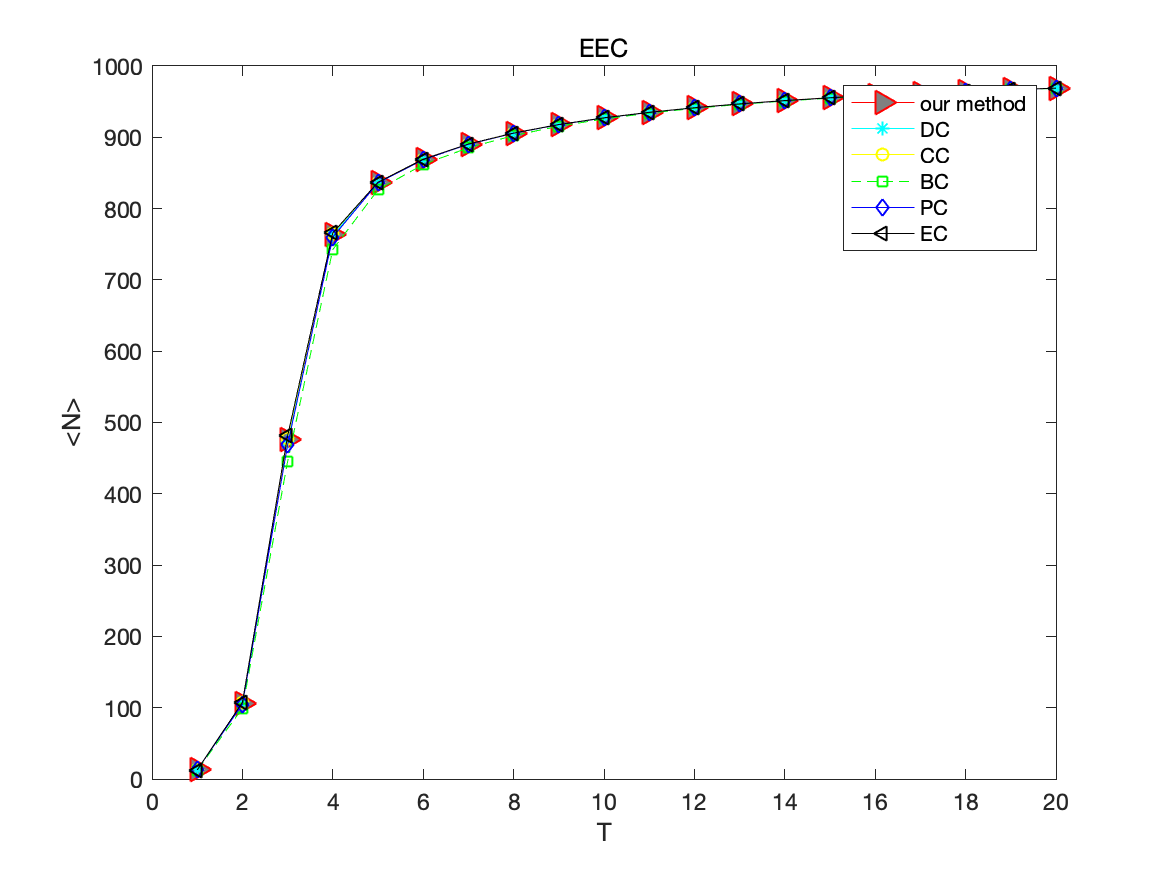}
\caption{The figure compares the infection ability of the top-100 nodes selected by different methods in EEC. All methods performance is similar while BC curve is sightly lower than other curves.}
\end{figure}
\begin{figure}[!htbp]
\centering
\includegraphics[width=0.8\textwidth]{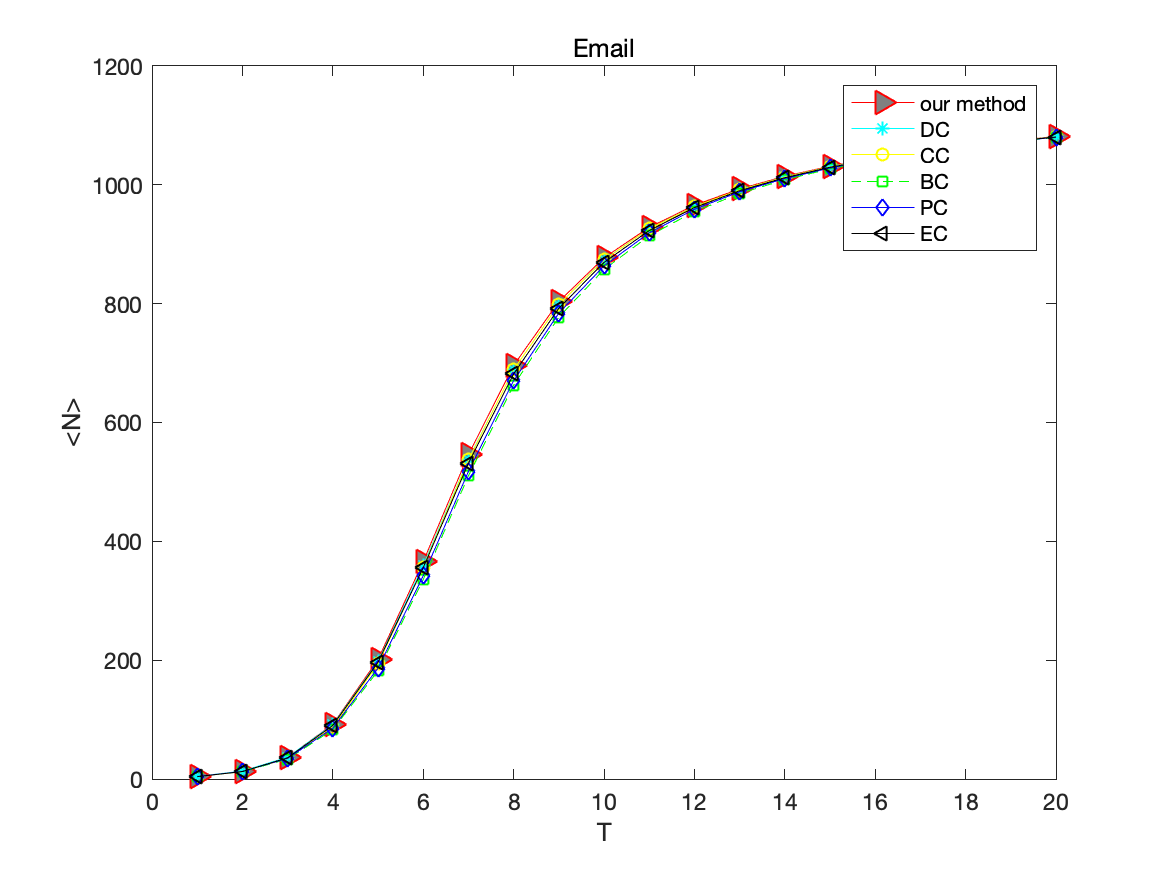}
\caption{The figure compares the infection ability of the top-100 nodes selected by different methods in Email. The performance of our method is better than others slightly. }
\end{figure}
\begin{figure}[!htbp]
\centering
\includegraphics[width=0.8\textwidth]{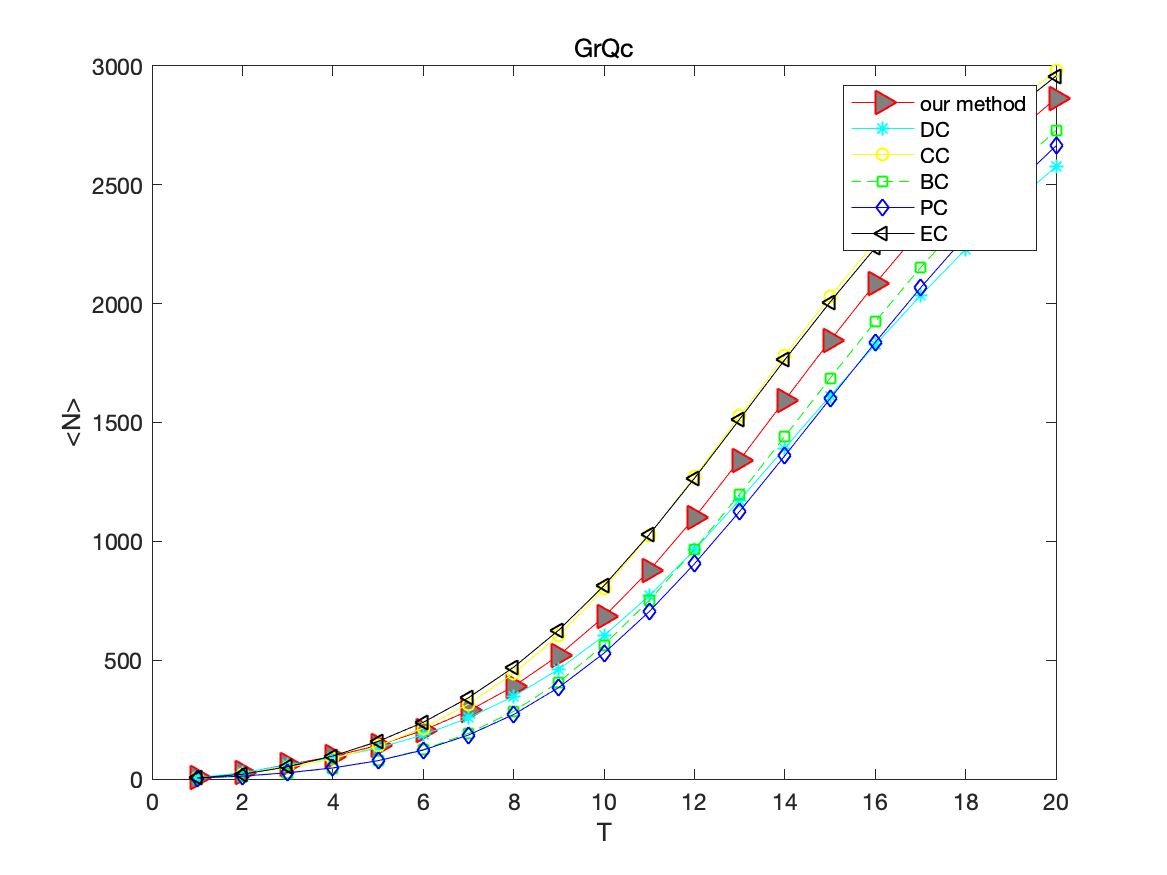}
\caption{The figure compares the infection ability of the top-100 nodes selected by different methods in GrQc. All methods performed different significantly. The top-100 nodes of CC and EC are the most influential, our method is the second.}
\end{figure}
\begin{figure}[!htbp]
\centering
\includegraphics[width=0.8\textwidth]{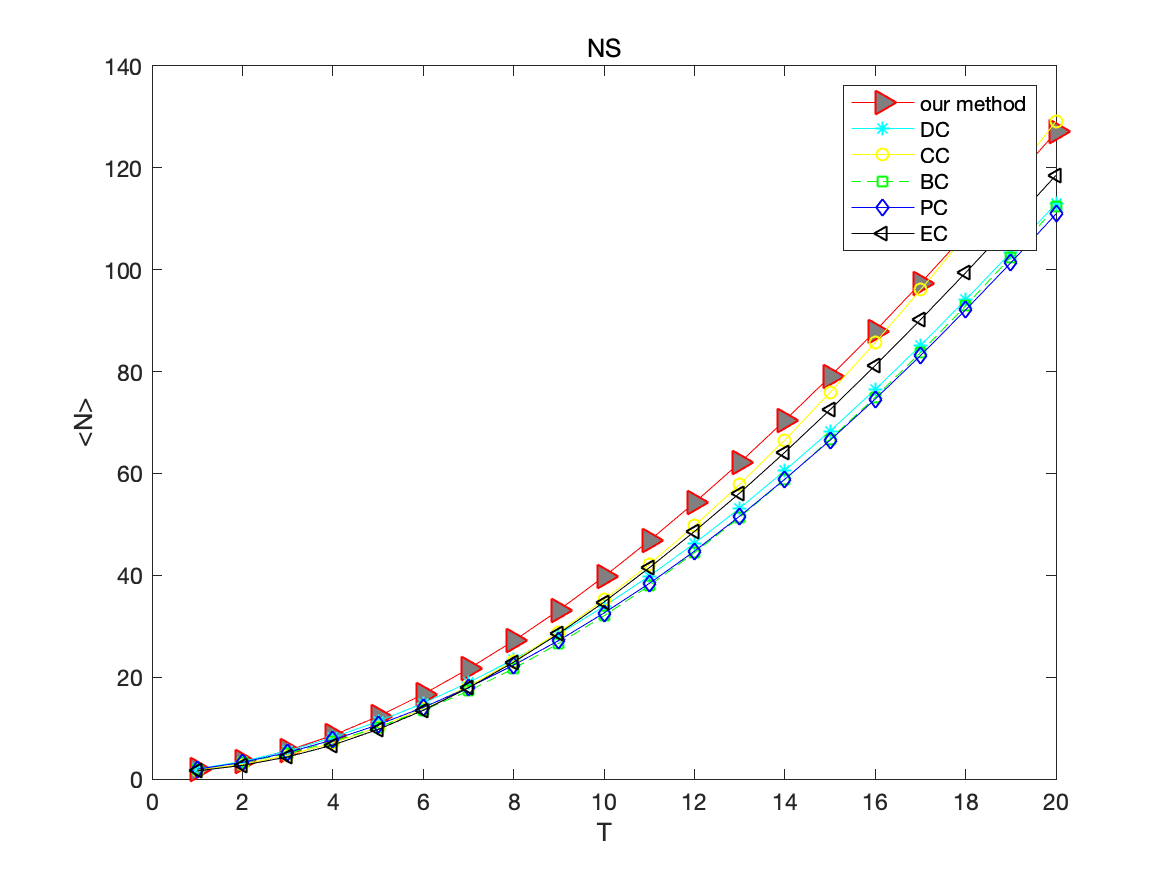}
\caption{The figure compares the infection ability of the top-100 nodes selected by different methods in NS. Our method’s curve rises fastest but PC’s curve rises slowly.}
\end{figure}
\begin{figure}[!htbp]
\centering
\includegraphics[width=0.8\textwidth]{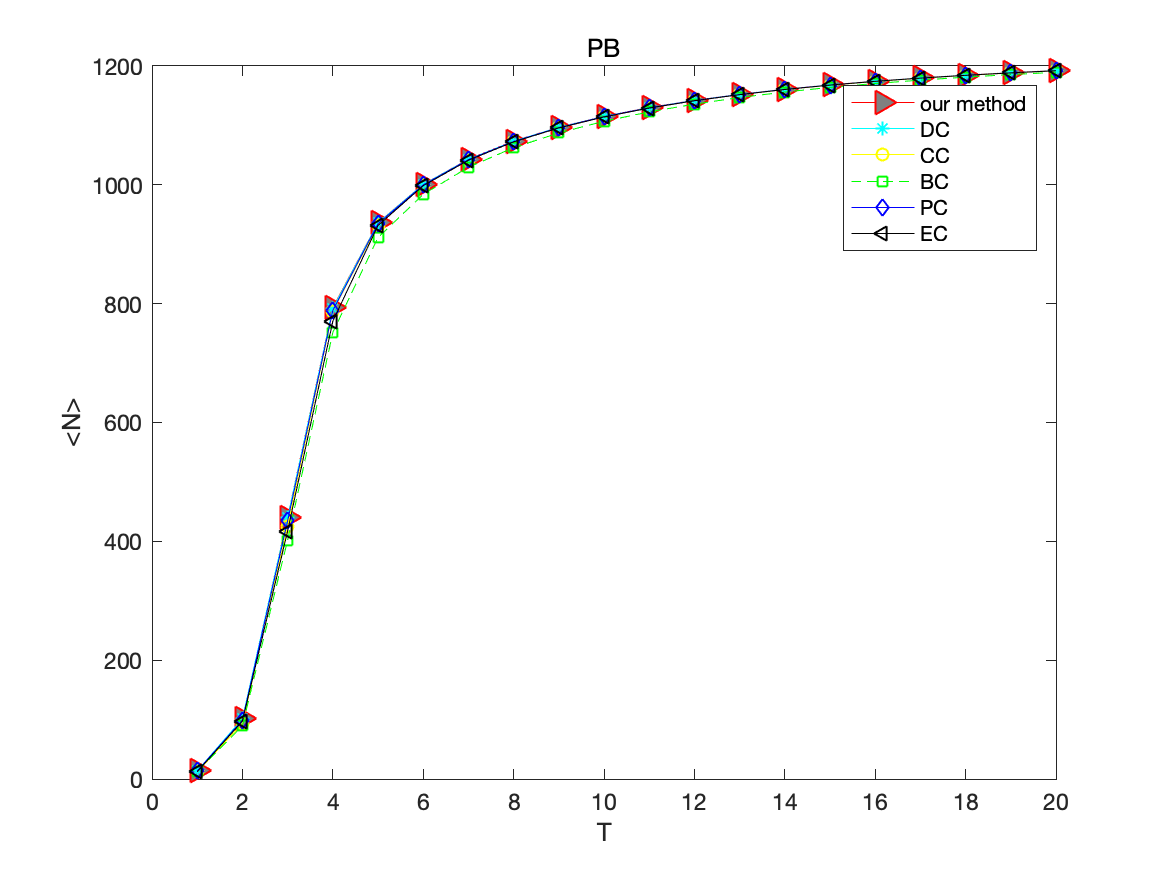}
\caption{The figure compares the infection ability of the top-100 nodes selected by different methods in PB. The difference among these methods are not obvious, which means they are basically consistent.}
\end{figure}
\begin{figure}[!htbp]
\centering
\includegraphics[width=0.8\textwidth]{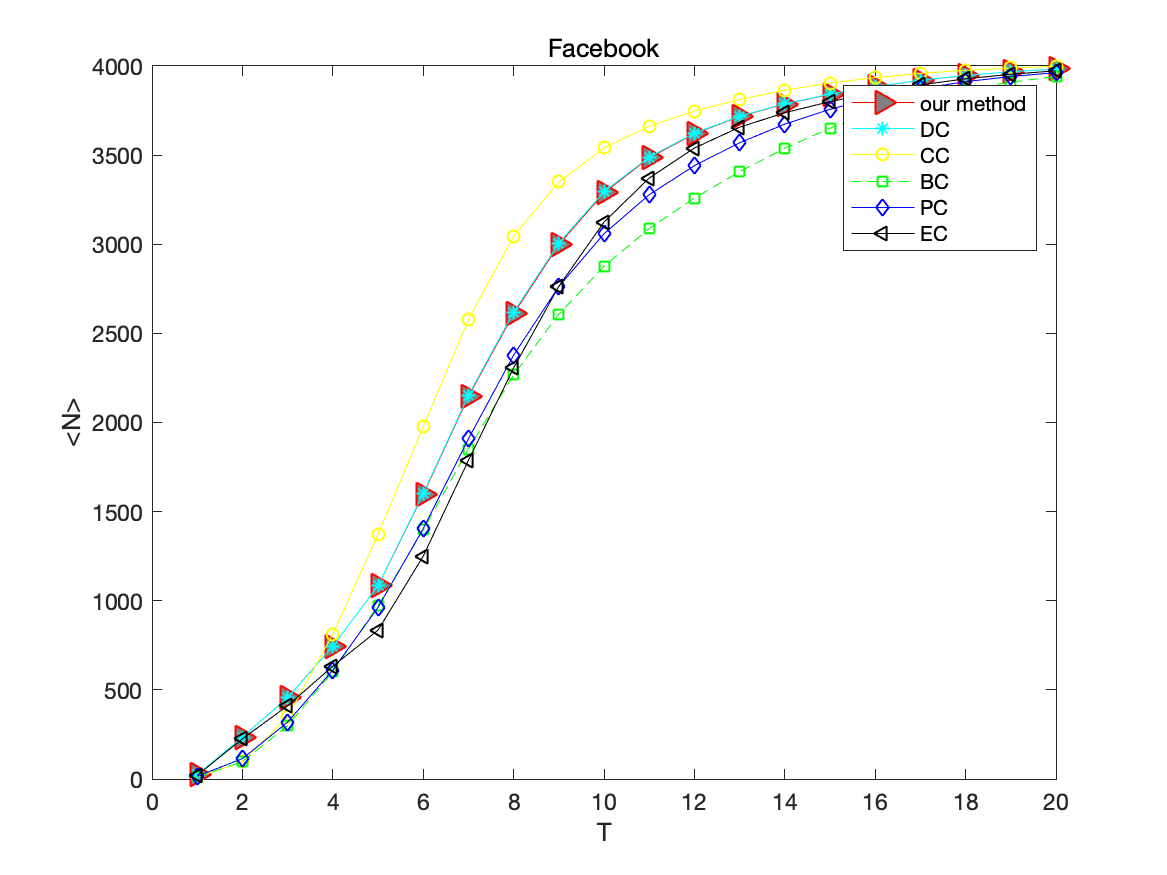}
\caption{The figure compares the infection ability of the top-100 nodes selected by different methods in Facebook. The trends of our method and DC are consistent. The CC curve rises fastest and BC rises slowly.}
\end{figure}
\begin{figure}[!htbp]
\centering
\includegraphics[width=0.8\textwidth]{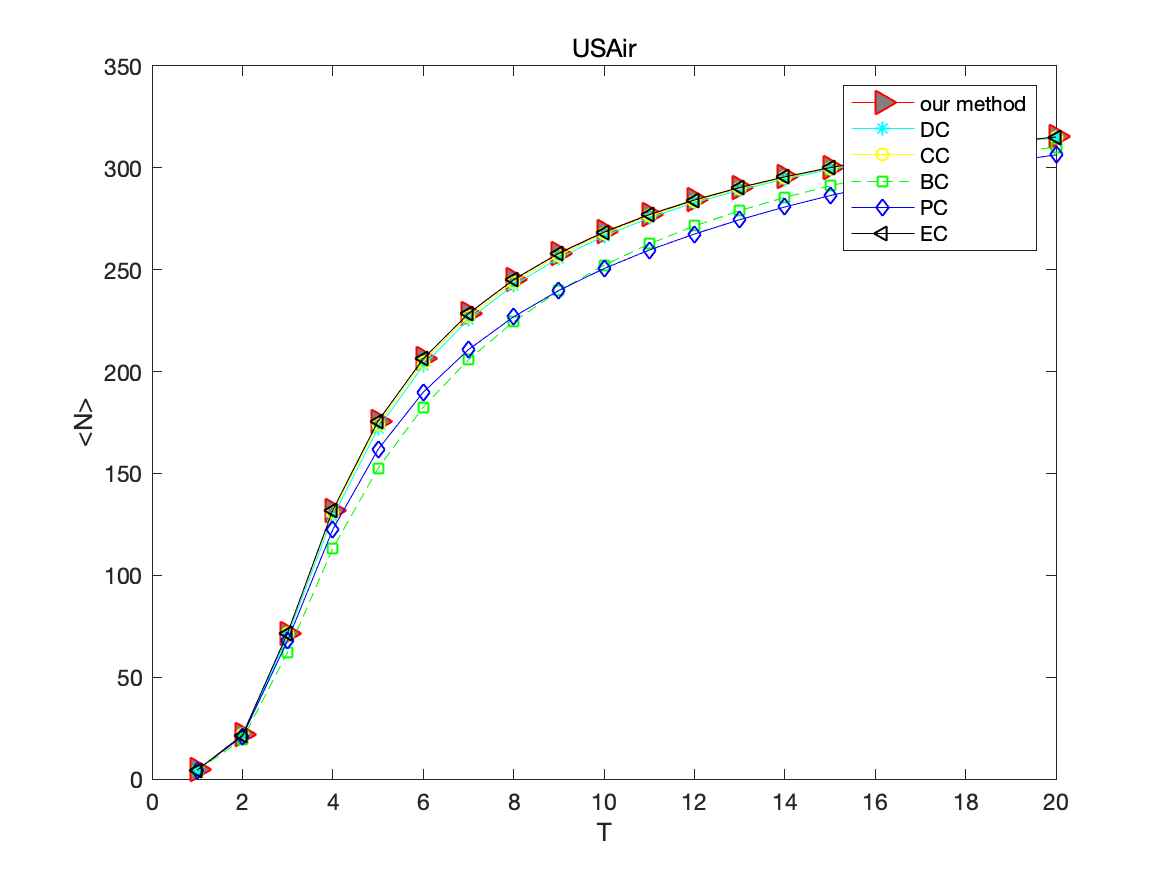}
\caption{The figure compares the infection ability of the top-100 nodes selected by different methods in USAir. Our method, EC, CC and DC performed similarly. The BC and PC rise more slowly than other methods.}
\end{figure}
\indent It can be seen that the curves corresponding to our proposed method and CC are always at the highest or second highest position. In addition, the slope of the curve corresponding to them is also very large in all networks mentioned, which means that the initial node set selected by the two has a stronger infection ability. That is to say, CC and our proposed method EffG can select influential nodes more accurately.%

\subsection{Comparison ranking results}
The top-20 vital nodes in the Jazz network ranked by different methods, our proposed method (EffG), DC, BC, CC, PC, EC, Gravity and SI model in which the $\beta$=0.2,t=20,N=50, are listed in Table \uppercase\expandafter{\romannumeral4}. The number of overlapping nodes in the set of the top-20 nodes sorted by our proposed method and the top-20 nodes set sorted by other methods are shown in Table \uppercase\expandafter{\romannumeral5}. The number of coincident nodes demonstrates the effectiveness of our method to a certain extent. As can be seen, in the Jazz network, the number of nodes that are consistent with the top-20 nodes obtained by our method and the top-20 nodes obtained by other methods are high. The high number of coincidences with other measures confirms the justifiability of our method and the unity with other methods.%
\begin{table*}[!htbp]
\centering
  \caption{Top-20 ranking of influential nodes in Jazz network by our proposed method (EffG), DC, BC, CC, PC, EC, Gravity and SI model.}
  \scalebox{1.4}{
  \begin{tabular}{ cccccccc}
  \hline
  our method &DC& CC& BC& PC& EC& SI& Gravity \\ 
  \hline
  8& 8& 8& 8& 8& 100& 8& 28 \\ 
  100& 100& 100& 155& 100& 4& 100&186\\
  4& 4& 131& 100& 131& 8& 4& 136\\
  131& 131& 194& 186& 4& 131& 194& 175\\
  194& 194& 69& 131& 186& 80& 53& 98\\
  80& 80& 4& 136& 136& 129& 131& 158\\
  69& 69& 53& 60& 69& 194& 111& 113\\
  162& 162& 111& 28& 28& 69& 133& 33\\
  53& 77& 162& 69&175& 53& 162& 23\\
  5& 5&129&175& 155& 32& 67& 9\\
  77& 53& 59& 194& 162& 84& 5& 87\\
  59& 32& 67& 9& 129& 85& 59& 86\\
  32& 59& 80& 32& 80& 130& 80& 4\\
  67& 186& 186& 111& 59& 162& 129& 77\\
  133& 67& 133& 4& 77& 77& 32& 131\\
  111& 133& 77& 59& 53& 133& 115& 38\\
  84& 28& 5& 79& 32& 115& 77& 178\\
  85& 111& 55& 113& 5& 89& 151& 72\\
  9& 9& 79& 151& 113& 59& 28& 16\\
  \hline
 \end{tabular}}
 \end{table*}
\begin{table*}[!htbp]
\centering
\caption{The number of same nodes between other methods and EffG.}
\scalebox{1.5}{
\begin{tabular}{ ccccccc}
  \hline
  DC& CC& BC& PC& EC& SI& Gravity\\
  \hline
  18& 17& 10& 14& 17& 17& 4\\
  \hline
  \end{tabular}}
  \end{table*}
\subsection{Relation of proposed method with other centrality methods}
The Kendall coefficient, $Kendall\ Tau$\cite{criado2013new}, is used to measure the correlation of two sequences. The absolute value of the Kendall coefficient is between 0 and 1. The larger the Kendall coefficient's absolute value, the stronger the correlation between the two sequences. If the Kendall coefficient between the two sequences is 0, it means the two sequences have no correlation. In this experiment, Kendall coefficient is used to measure the correlation between sequences generated by different identification methods and the sequence generated by the SI model, thereby inferring the effectiveness of the identification method. The greater the absolute value of the Kendall coefficient is, the more valid the identification method is.\\
\indent Given two sequences with $N$ elements, $X=(x_1,x_2,x_3,…,x_n)$ and $Y=(y_1,y_2,y_3,…,y_n)$. Let $(x_i,y_i)$ by a set of sequence pairs. For any pairs $(x_i,y_i)$ and $(x_j,y_j)$ that $i \ne j$, if both $x_i>x_j$ and $y_i>y_j$ or both $x_i<x_j$ and $y_i<y_j
$, they are classified as concordant sequence pairs. While if both $x_i>x_j$ and $y_i<y_j$ or both $x_i<x_j$ and $y_i>y_j$, they are classified as the discordant sequence pairs. The Kendall’s Tau of two sequences $X$ and $Y$, is defined as follows.
\begin{equation}
tau=\frac{n_+ - n_-}{N\times (N-1)}
\end{equation}
Where $n_+$ and $n_-$ are the number of concordant sequence pairs and discordant sequence pairs respectively, $N$ is the total number of sequence pairs.\\
\indent In this experiment, the evaluation of the effectiveness of the method is based on the correlation with the SI model. In all data sets, the different infection probability $\beta$ was given to the SI model respectively to obtain a standard centralized sequence. Then $Kendall's Tau$ of SI model sequence and other method’s sequence was calculated. In the experiment, the infection probability $\beta$ changed from 0.2 to 1.6, and the SI model was independently work 50 times with the infection time $t = 5$ to take the average on different networks. The experimental results are shown in Figs.17-22, where $tau$ represents the value of $Kendall's Tau$. Higher tau value indicates stronger positive correlation between centrality method and SI model.\\
\indent As can be seen that CC has the strongest correlation with SI and  CC is gradually close to SI as $\beta$ increases, which means that it has a strong positive correlation with the SI model. Our proposed method performs well in general although the correlation is a little weak on some networks. For instance, the $tau$ of our method is the second highest when $\beta>0.8$ in Figs.20-21.
\begin{figure}[!htbp]
\centering
\includegraphics[width=0.8\textwidth]{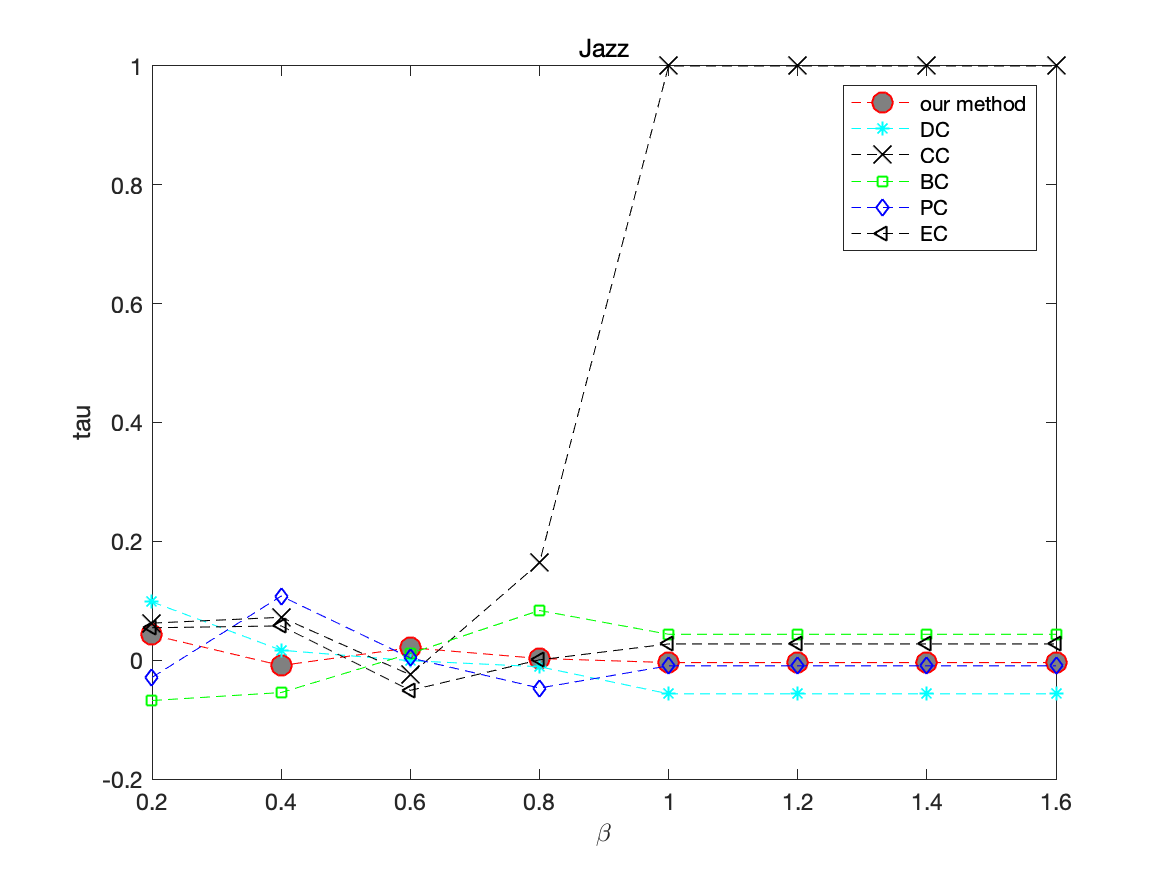}
\caption{The figure compares the tau of SI model sequence and other method’s sequence in Jazz. CC has the strongest correlation with SI and CC is gradually close to SI as $\beta$ increases.}
\end{figure}
\begin{figure}[!htbp]
\centering
\includegraphics[width=0.8\textwidth]{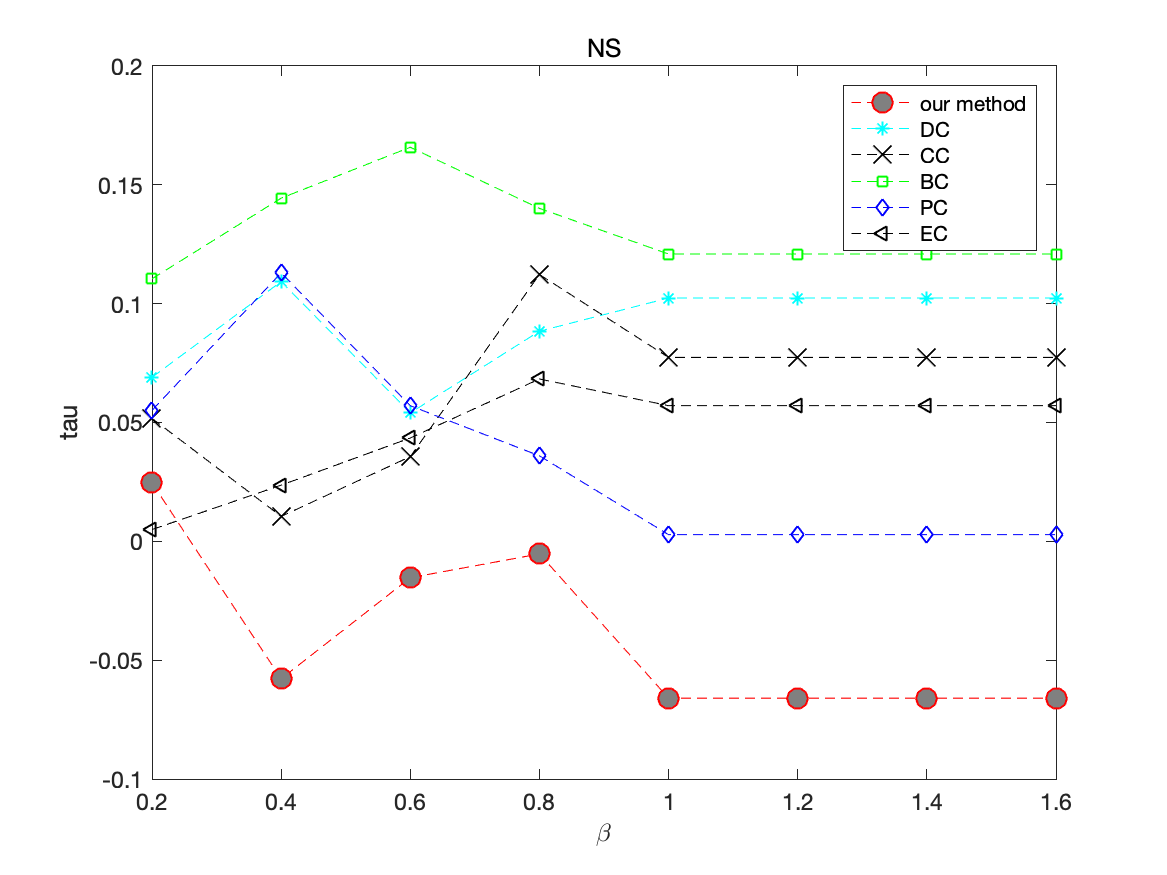}
\caption{The figure compares the tau of SI model sequence and other method’s sequence in NS. The tau of BC is the highest while the tau of our method is the lowest when $\beta>0.24$.}
\end{figure}
\begin{figure}[!htbp]
\centering
\includegraphics[width=0.8\textwidth]{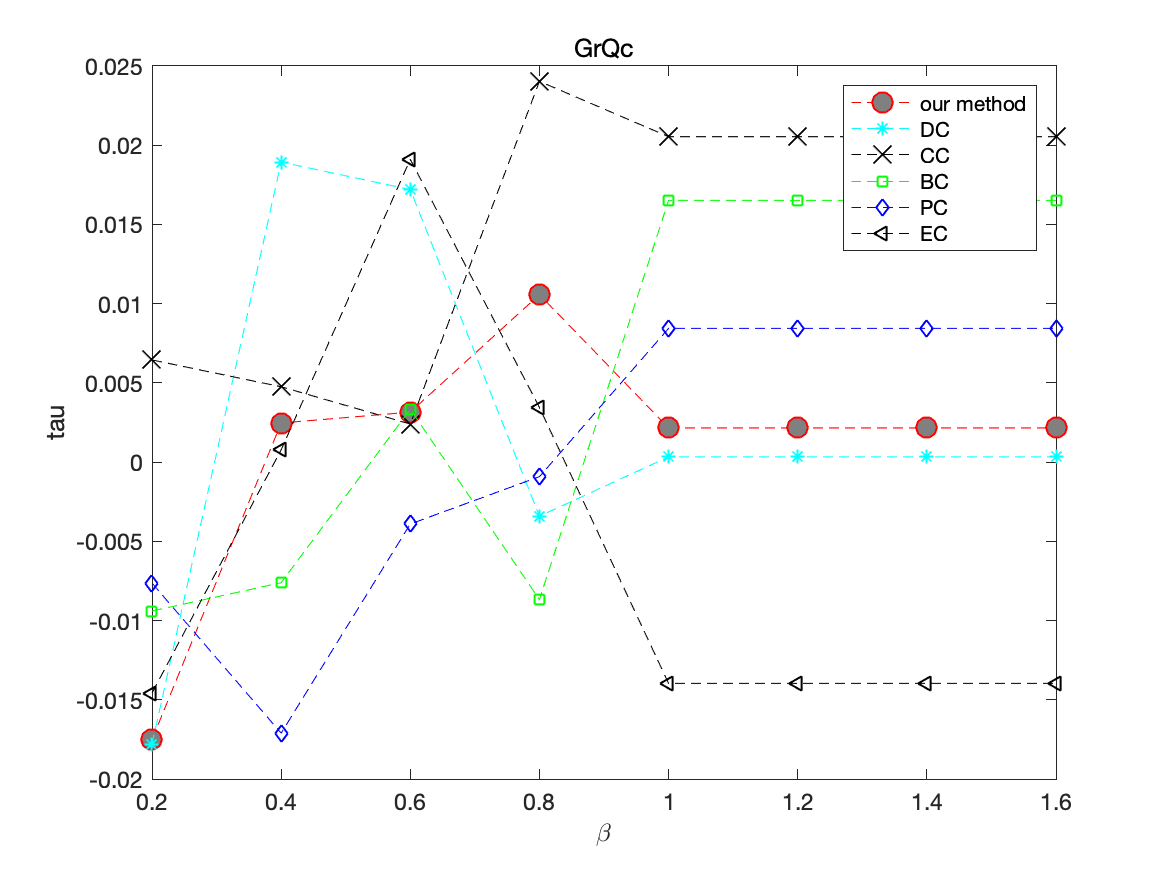}
\caption{The figure compares the tau of SI model sequence and other method’s sequence in GrQc. The tau of our method is in the middle basically, and as $\beta$ increases, all methods have large fluctuations in tau value.}
\end{figure}
\begin{figure}[!htbp]
\centering
\includegraphics[width=0.8\textwidth]{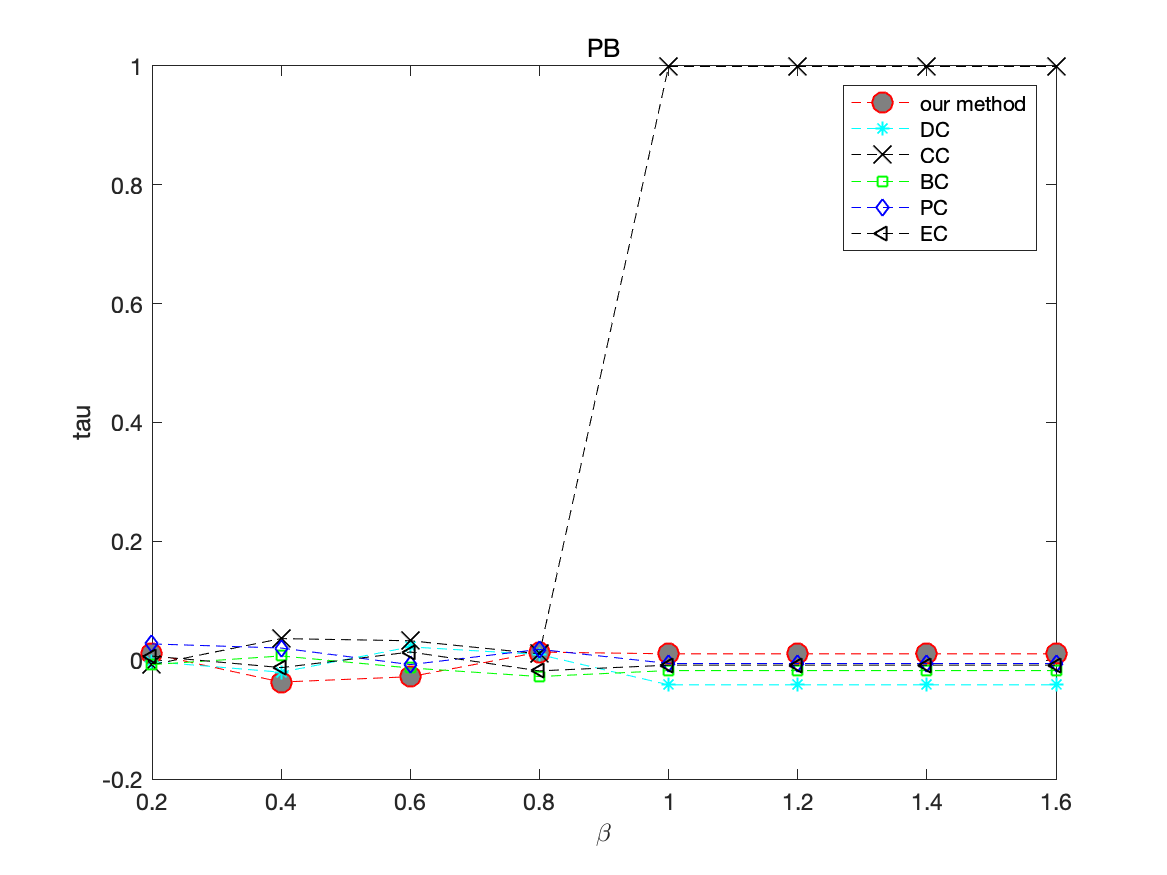}
\caption{The figure compares the tau of SI model sequence and other method’s sequence in PB. The tau of our method is highest when $\beta>0.8$ while when $\beta<0.6$ it is lowest. And tau of DC is lowest when $\beta>0.9$.}
\end{figure}
\begin{figure}[!htbp]
\centering
\includegraphics[width=0.8\textwidth]{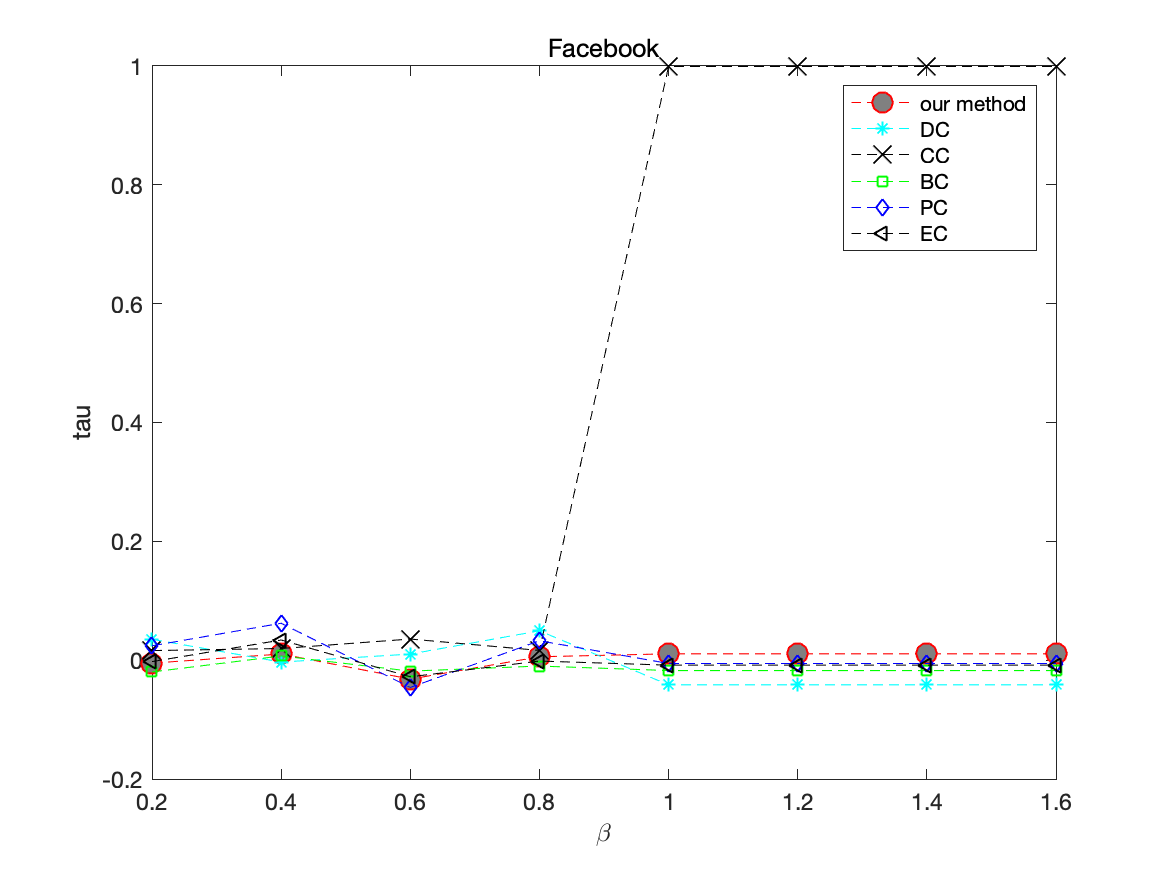}
\caption{The figure compares the tau of SI model sequence and other method’s sequence in Facebook. The tau of our method, PC and EC are the lowest when $\beta=0.6$. As $\beta$ changes, CC and tau are gradually close to the SI and the tau of our method is the second highest when $\beta>0.9$.}
\end{figure}
\begin{figure}[!htbp]
\centering
\includegraphics[width=0.8\textwidth]{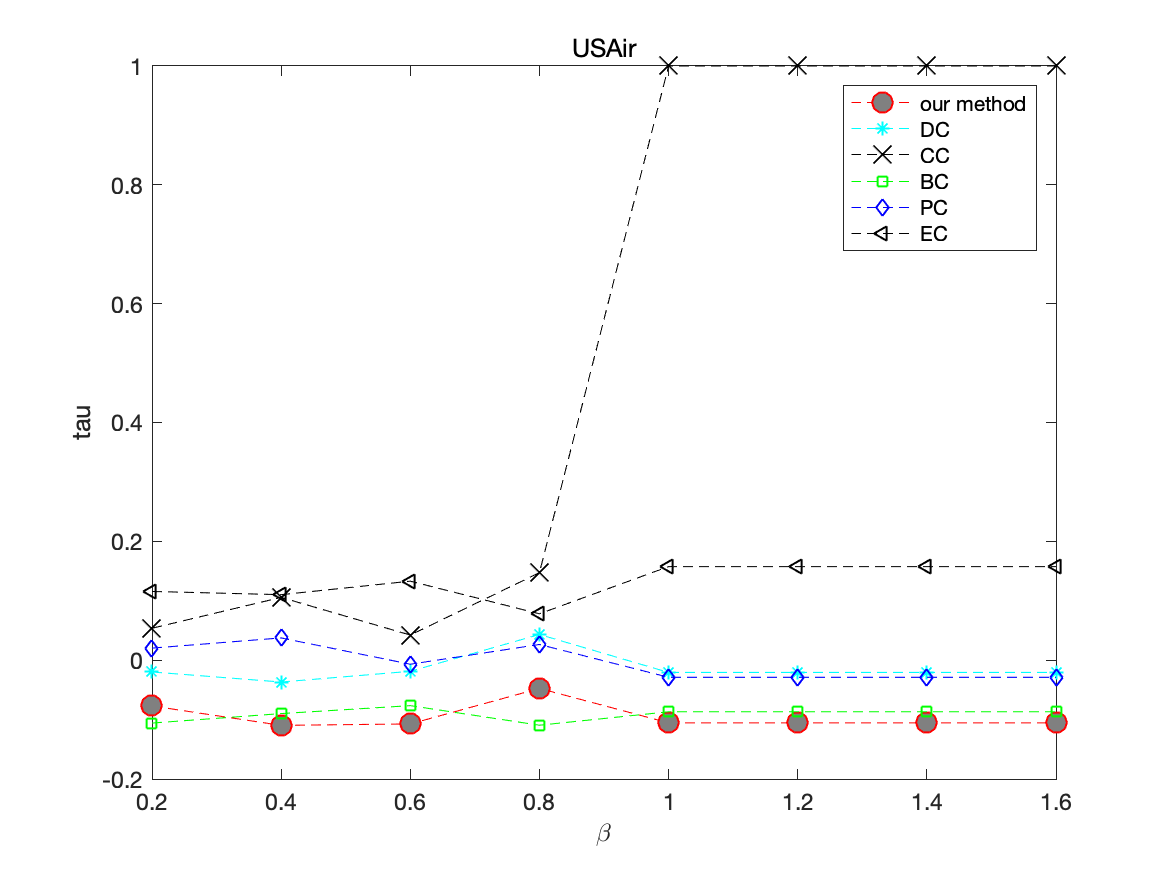}
\caption{The figure compares the tau of SI model sequence and other method’s sequence in USAir. The tau of our method and BC are lower.}
\end{figure}
\subsection{Compare the correlations between proposed method and SI model.}
In this experiment, four real-world networks were used to evaluate the feasibility of our method, including Jazz, NS, Email and USAir. First, the ranking of nodes on each network is derived by different methods, DC, BC, CC, EC and our proposed method. Then, each node will be used as the initial infected node in the SI model, and the final number of infected nodes will be calculated by t = 20. Finally, the correlation between the node ranking and the final number of nodes infected by them, denoted as $<N>$, will be established. The results are shown in Figs.23-26.\\
\indent The node with higher ranking is of stronger capability to infect other nodes, which means the node is more influential. That is to say, the higher the ranking of a node is, the greater the number of nodes eventually infected by it should be. The lower the node ranking is, the smaller the final number of infected nodes should be. Hence, the curve corresponding to a good identification method should basically continue to decline. As can be seen in Figs.23-26, the curve corresponding to our method is continuously declining, and has little fluctuation compared with other methods. However, it can be easily found that the curve corresponding to BC fluctuates greatly during the decline and does not clearly show a downward trend. Therefore, it can be inferred that our proposed method EffG is valid and reasonable compared to other methods to some extent.%
\begin{figure}[!htbp]
\centering
\includegraphics[width=0.8\textwidth]{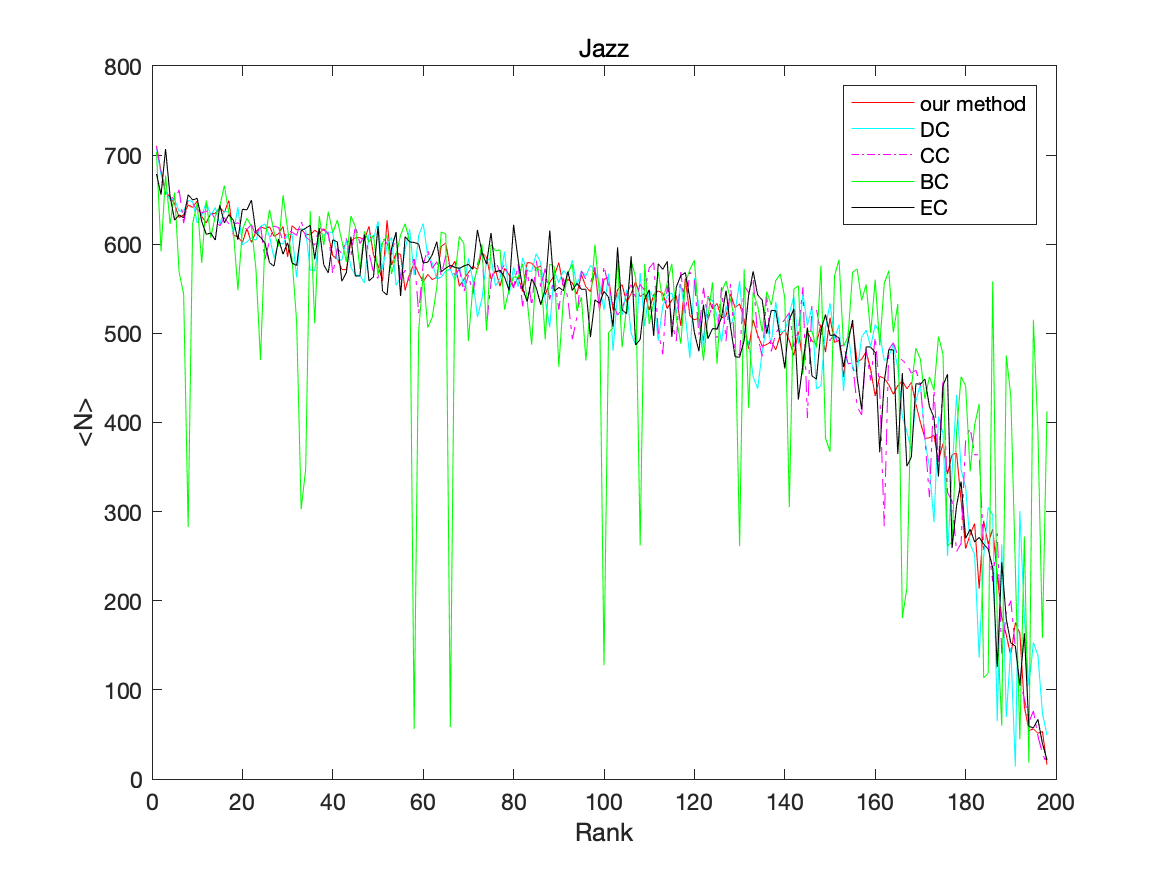}
\caption{The figure describes the correlation between different methods and SI model in Jazz. As the ranking of nodes changes, the N value of BC fluctuates greatly. And our method continues to decline and is most stable}
\end{figure}
\begin{figure}[!htbp]
\centering
\includegraphics[width=0.8\textwidth]{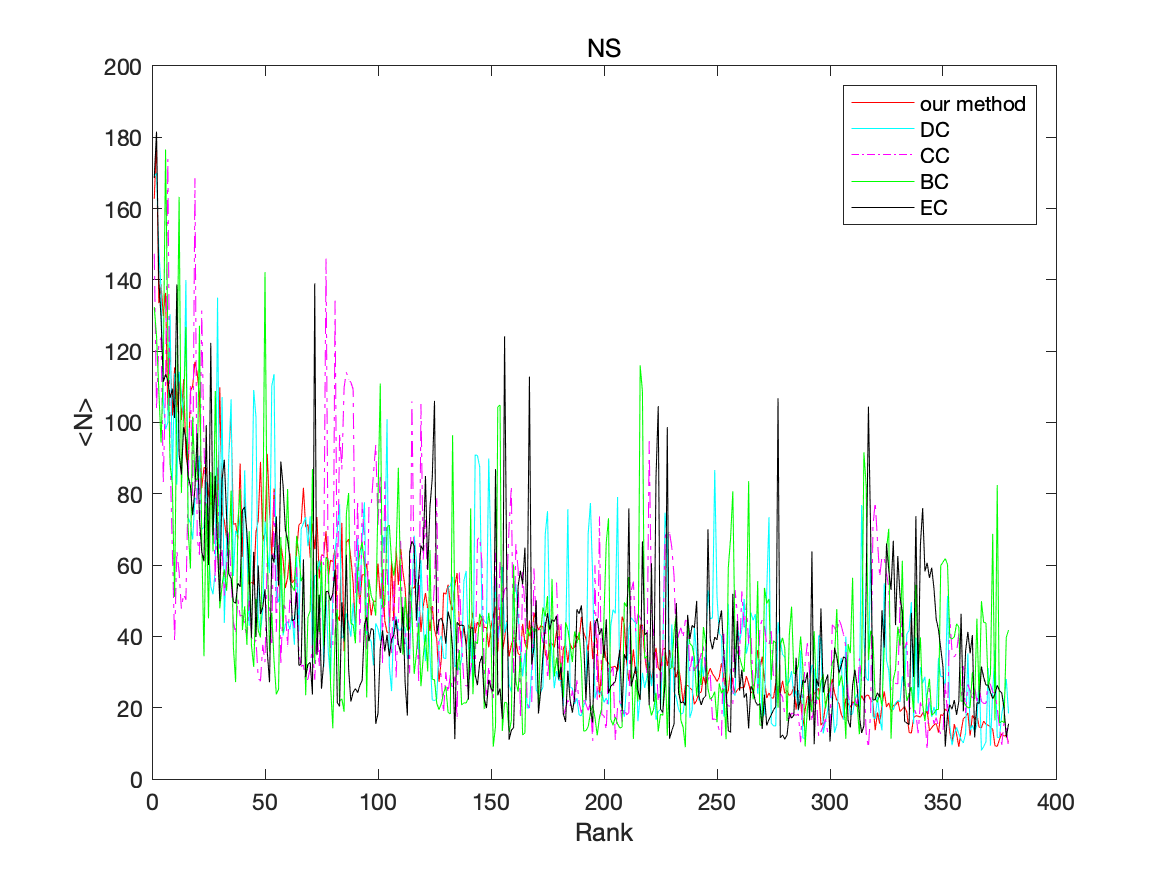}
\caption{The figure describes the correlation between different methods and SI model in NS. The curves of CC, BC, EC and DC fluctuates greatly. While our method continues to decline and is stable. }
\end{figure}
\begin{figure}[!htbp]
\centering
\includegraphics[width=0.8\textwidth]{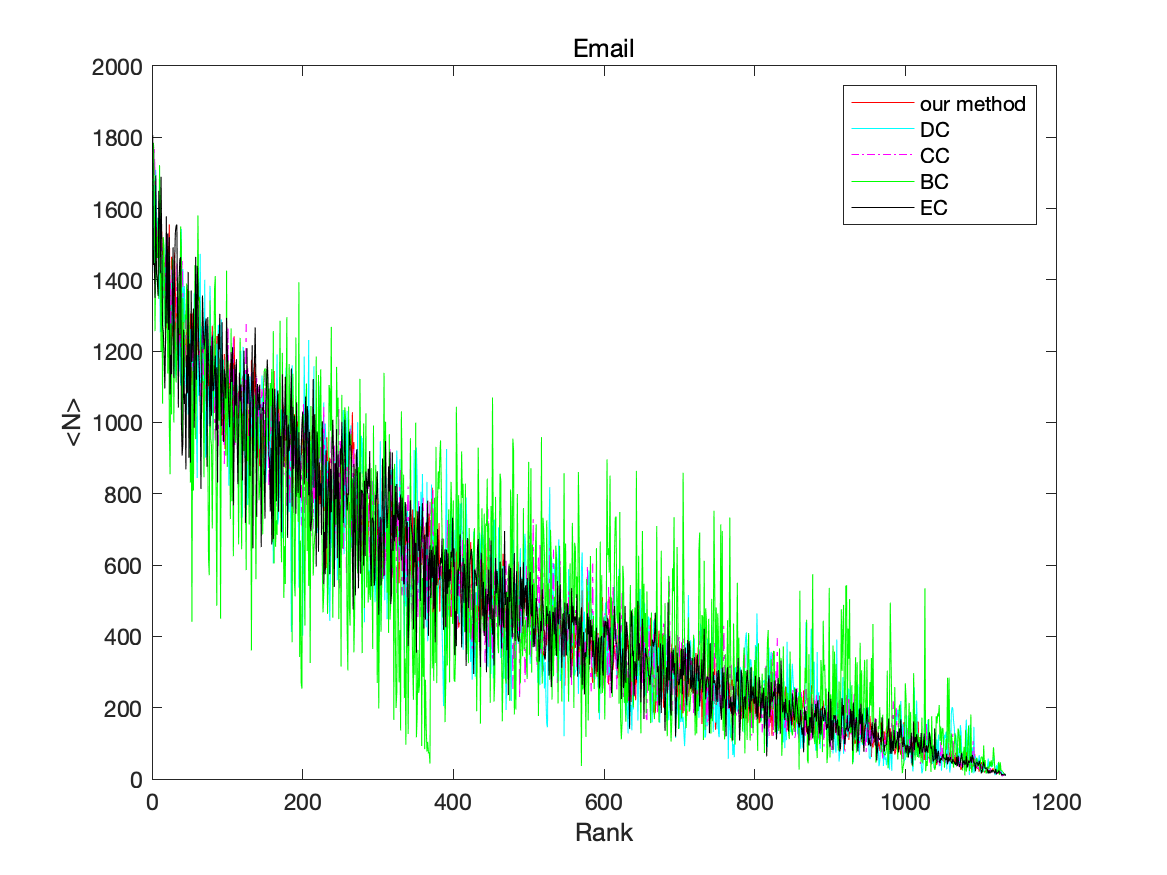}
\caption{The figure describes the correlation between different methods and SI model in Email. Overall trend of these curves is declining. And the curve of BC fluctuates the most greatly. While our method and CC are more stable.}
\end{figure}
\begin{figure}[!htbp]
\centering
\includegraphics[width=0.8\textwidth]{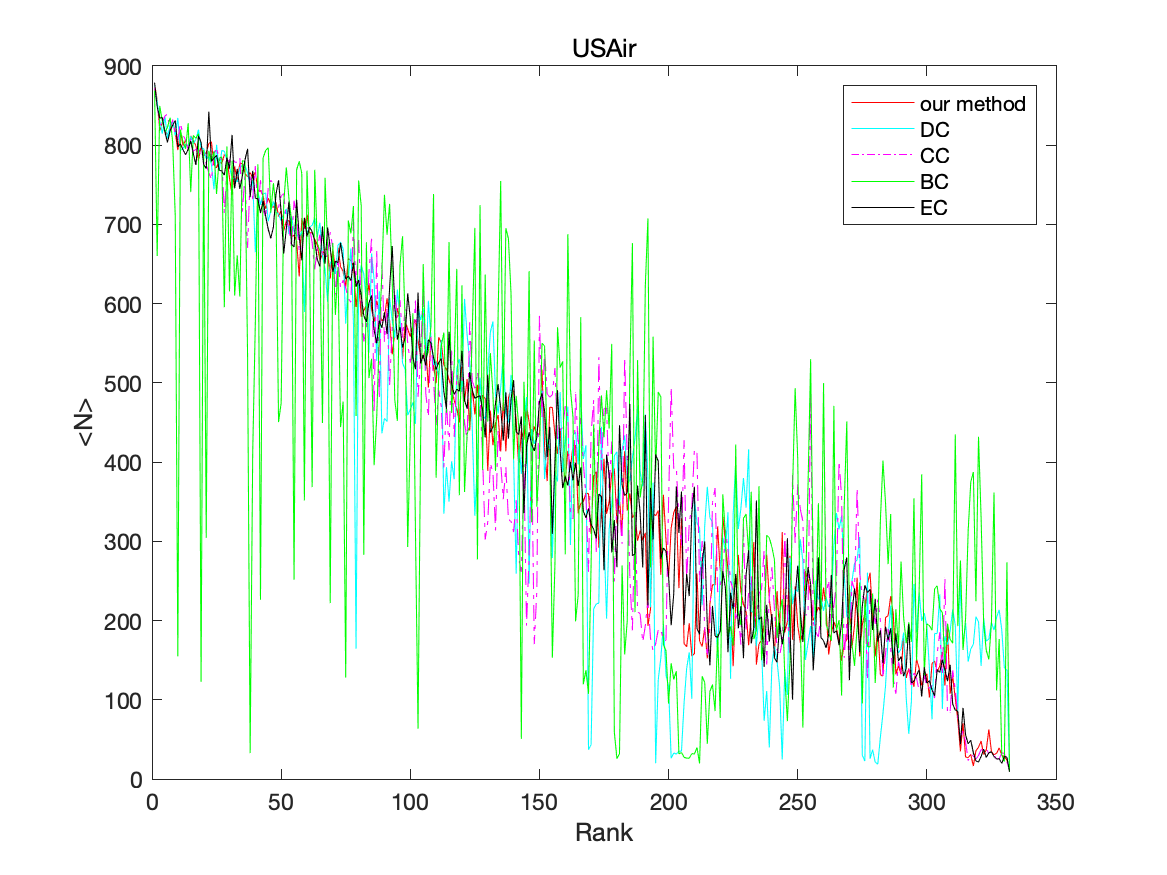}
\caption{The figure describes the correlation between different methods and SI model in USAir. Overall trend of these curves is declining. The curve of BC fluctuates greatly. While our method and EC are more stable than the others}
\end{figure}

\section{Conclusion}
In this paper, an original and novel method for identifying the influence node is proposed: an effective distance gravity model. Instead of considering single-dimensional factors, our proposed EffG comprehensively considers the local information of the node and global information of the network based on the idea of multi-source information fusion. An important contribution is that the EffG uses the effective distance to replace traditional static Euclidean Distance. EffG is able to take full advantage of the dynamic information exchange between nodes in the real-world network. In addition, EffG can help us unravel the topology of the network that drives many dynamic information propagation processes. Importantly, the identification of influential nodes by EffG is aligned to real-world conditions. In order to verify the effectiveness and feasibility of this method, a variety of experiments were conducted on eight real-world networks and compared with six existing well-known methods. The experimental results indicated that our method performs well under dynamic information propagation and across several test-examples, thereby demonstrating its potential applications in network science, biological and social system, time series and information propagation.

\section*{Conflict of interests}
The authors declare that there is no conflict of interests regarding the publication of this paper.
\section*{Data Availability Statements}
The data that support the findings of this study are available from the corresponding author upon reasonable request.
\section*{Acknowledgment}
The work is partially supported by National Natural Science Foundation of China (Grant Nos. 61573290, 61503237).

\bibliographystyle{elsarticle-num}
\bibliography{reff}

\end{document}